\documentclass[lettersize,journal]{IEEEtran}
\usepackage{amsmath,amsfonts}
\usepackage{algorithm}
\usepackage{array}
\usepackage[caption=false,font=normalsize,labelfont=sf,textfont=sf]{subfig}
\usepackage{textcomp}
\usepackage{stfloats}
\usepackage{url}
\usepackage{verbatim}
\usepackage{graphicx}
\usepackage{cite}

\usepackage{booktabs}
\usepackage{capt-of}

\usepackage{listings}		
\usepackage{enumerate}	

\usepackage{hyperref}

\newcommand{\envleftshift}{}
\renewcommand{\envleftshift}{0in} 	

\newtheorem{definition}{\hspace*{-\envleftshift}Definition}[section]

\newtheorem{lemma}{\hspace*{-\envleftshift}Lemma}[section]
\newtheorem{proposition}{\hspace*{-\envleftshift}Proposition}[section]
\newtheorem{theorem}{\hspace*{-\envleftshift}Theorem}[section]

\newtheorem{remark}{\hspace*{-\envleftshift}Remark}[section]

\usepackage{color,soul}

\usepackage[table]{xcolor}

\usepackage{gensymb}
\usepackage{amssymb}

\usepackage{mathtools}
\usepackage{mathrsfs}

\usepackage{tensor}

\newcommand{\norm}[1]{\left\Vert #1\right\Vert}

\usepackage{float}
\usepackage{placeins} 

\usepackage{hhline} 		
\usepackage{multirow}		
\usepackage{chngpage}
\usepackage{lscape} 		

\usepackage{tablefootnote} 	

\usepackage{tikz}

\usepackage{algpseudocode}

\usepackage{mdframed}

\newcommand{\relpath}{}
\newcommand{\relpathone}{}
\newcommand{\relpathtwo}{}

\newdimen\figsize
\newcommand{\lightshade}{20}
\newcommand{\darkshade}{50}

\newcommand{\Hinf}{{\cal H}^\infty}

\usepackage{bibentry}
\nobibliography*

\newcommand{\bibpath}{}

\newcommand\copyrighttext{%
	\footnotesize \textcopyright 2023 IEEE. Personal use of this material is permitted. Permission from IEEE must be obtained for all other uses, in any current or future media, including reprinting/republishing this material for advertising or promotional purposes, creating new collective works, for resale or redistribution to servers or lists, or reuse of any copyrighted component of this work in other works.
  }
\newcommand\copyrightnotice{%
\begin{tikzpicture}[remember picture,overlay]
\node[anchor=south,yshift=10pt] at (current page.south) {\fbox{\parbox{\dimexpr\textwidth-\fboxsep-\fboxrule\relax}{\copyrighttext}}};
\end{tikzpicture}%
}

\newcommand{\CTRLSecCSI}{IX-A}

\newcommand{\CTRLFigIRLWeights}{4c}

\begin{document}

%
%


%
%

%
%

\title{Continuous-Time Reinforcement Learning: New Design Algorithms with Theoretical Insights and Performance Guarantees}

\author{

    Brent A. Wallace and Jennie Si,~\IEEEmembership{Fellow,~IEEE}

    \thanks{This work was supported in part by the NSF under Grants 1808752 and 2211740. The work of Brent A. Wallace was supported in part by the NSF under Graduate Research Fellowship Grant 026257-001.}
    \thanks{Brent A. Wallace and Jennie Si are with the Department of Electrical, Computer \& Energy Engineering, Arizona State University, Tempe, AZ 85287 USA (e-mail: bawalla2@asu.edu; si@asu.edu).}

}   

%
%


%
%


\maketitle

%
%

\copyrightnotice



%
%


%
%

\begin{abstract}

Continuous-time nonlinear optimal control problems hold great promise in real-world applications. After decades of development, reinforcement learning (RL) has achieved some of the greatest successes as a general nonlinear control design method. However, a recent comprehensive analysis of state-of-the-art continuous-time RL (CT-RL) methods, namely, adaptive dynamic programming (ADP)-based CT-RL algorithms, reveals they face significant design challenges due to their complexity, numerical conditioning, and dimensional scaling issues. Despite advanced theoretical results, existing ADP CT-RL synthesis methods are inadequate in solving even small, academic problems. The goal of this work is thus to introduce a suite of new CT-RL algorithms for control of affine nonlinear systems. Our design approach relies on two important factors. First, our methods are applicable to physical systems that can be partitioned into smaller subproblems. This constructive consideration results in reduced dimensionality and greatly improved intuitiveness of design. Second, we introduce a new excitation framework to improve persistence of excitation (PE) and numerical conditioning performance via classical input/output insights. 
Such a design-centric approach is the first of its kind in the ADP CT-RL community.
In this paper, we progressively introduce a suite of (decentralized) excitable integral reinforcement
learning (EIRL) algorithms. We provide convergence and closed-loop stability guarantees, and we demonstrate these guarantees on a significant application problem of controlling an unstable, nonminimum phase hypersonic vehicle (HSV).

\end{abstract}




%
%


%
%

\section{Introduction \& Motivation}\label{sec:introduction}

%
%

\IEEEPARstart{T}{he} origins of modern approaches to optimal control problems are rooted in the 1960s with the inception of dynamic programming (DP) by Bellman \cite{Bellman_DP:book}.
%
%
Reinforcement learning (RL) emerged as a systematic method in the early 1980s \cite{Barto_Sutton_Anderson_RL:1983,Sutton_Barto_RL_opt_contr:book}
with the potential to combat the curse of dimensionality in DP. 
%
An important branch of research work on RL for decision and control is covered under the scope of ADP \cite{Werbos_RL_contr:book:1991, Werbos_ADP:incollection:1992, Si_Barto_Powell_Wunsch_approx_DP:book, Bertsekas_opt_contr:book}, which uses approximation and learning to solve the optimal control problem for both continuous-time (CT) and discrete-time (DT) dynamical systems.

%
%


%
%

\noindent\textbf{DT-RL Theoretical/Application Successes.} DT-RL algorithms (cf. \cite{Lewis_Vrabie_Vamvoudakis_RL_Contr_Overview:2012,Kiumarsi_KG_Vamvoudakis_H_Modares_FL_Lewis_RL_survey:2018} for review) have demonstrated excellent stability, convergence, and approximation guarantees. Representative theoretical works include 
\cite{Mu_Wang_He_NDP:2017,Liu_Wei_PI_DT:2015,Wei_Liu_Lin_Song_PI_ADP_DT:2016,Guo_J_si_Liu_Mei_PI_approx:2018,Liu_Wei_Yan_GPI_ADP_DT:2015,Gao_Si_Wen_Li_Huang_prosthetics:2021} 
that are based on the policy iteration (PI) framework, and 
\cite{Al-Tamimi_FL_Lewis_Abu-Khalaf_VI_proof:2008,Wang_Liu_Wei_Zhao_Jin_ADP_DT:2012,Liu_Sun_J_Si_Guo_Mei_dHDP_boundedness_result:2012,Liu_Wei_VI_DT_approx_error:2013,Wei_Wang_Liu_Yang_finite_approx_DT:2014} 
that are based on the value iteration (VI) framework. 
%
DT-RL methods have demonstrated great successes in addressing complex continuous state and control problems, including
energy-efficient data centers \cite{Farahnakian_Lijeberg_Plosila_energy_efficient_data_center:2014}, aggressive ground robot position control \cite{Mondal_AA_Rodriguez_Manne_Das_BA_Wallace_wheeled_robot:2019,Mondal_BA_Wallace_AA_Rodriguez_wheeled_robot:2020}, power system stability enhancement \cite{Lu_J_Si_Xie_dHDP_power_sys:2008,Guo_Liu_J_Si_He_Harley_Mei_power_sys_stability:2015,Guo_Liu_J_Si_He_Harley_Mei_power_sys_freq_contr:2016}, industrial process control \cite{Wei_Liu_water_gas_process_contr:2014,Jiang_Fan_Chai_FL_Lewis_flotation_process_contr:2019}, Apache helicopter stabilization, tracking, and reconfiguration control \cite{Enns_J_Si_Apache_NDP:2002,Enns_Si_helicopter:2003,Enns_J_Si_helicopter_rotor_failure:2003}, waste water treatment \cite{Yang_Cao_Meng_J_Si_wastewater:2021}, and wearable robots to enable continuous and stable walking \cite{Wen_Liu_J_Si_Huang_ADP_transfemoral:2016,Wen_J_Si_Gao_Huang_Huang_prosthesis_lower_limb_ADP:2017,Wen_J_Si_Brandt_Gao_Wang_prosthesis_RL_contr:2019,Wu_Li_Yao_Liu_J_Si_Huang_prosthesis_RL_impedance_contr:2022,Li_Wen_Gao_J_Si_Huang_prosthesis_personalized_impedance:2022,Wu_Zhong_BA_Wallace_Gao_Huang_J_Si_prosthesis_symmetry:2022}.

%
%

\IEEEpubidadjcol

%
%

\noindent\textbf{CT-RL Limitations and Lack of Real-World Synthesis Capability.} On the other hand, CT-RL algorithms, which center around representative and seminal works of \cite{Vrabie_Lewis_IRL:2009,Vamvoudakis_Lewis_SPI:2010,Jiang_ZP_Jiang_Robust_ADP:2014,Bian_ZP_Jiang_ADP_VI:2021}, have seen fewer theoretical developments when compared to their DT-RL counterparts. The first-of-its-kind comprehensive CT-RL numerical analysis \cite{BA_Wallace_J_Si_CT_RL_review:2022} reveals that prevailing CT-RL algorithms suffer acutely from the following major challenges:
\begin{enumerate}[1)]

	\item\textbf{Curses of Conditioning, Dimensionality: \cite[Sections IX-A, IX-C]{BA_Wallace_J_Si_CT_RL_review:2022}.}  Systematically achieving PE proves difficult, even for low-order systems with a small number of
    neural network (NN) basis functions. As a result, CT-RL algorithms uniquely suffer from intrinsic numerical conditioning issues. Conditioning plays a vital role in quantifying the sensitivity of a regression (i.e., solving for the NN weights) to perturbations in the regression data (due to, e.g., model uncertainty, measurement noise, etc.) \cite{Higham_numerical_algs:book}. The studies in \cite{BA_Wallace_J_Si_CT_RL_review:2022} show that conditioning of existing CT-RL algorithms is oftentimes high enough to cause NN weight divergence and complete numerical breakdowns. 
    Dimensional scalability issues of CT-RL algorithms are also exceptionally acute. Conditioning is shown to increase by multiple orders of magnitude in response to small increments in network basis dimension, limiting real-world applicability.	

	\item\textbf{CT-RL Algorithm Complexity Issues: \cite[Section IX-B]{BA_Wallace_J_Si_CT_RL_review:2022}.}  Underlying algorithm complexity creates large discrepancies between theoretical results and 
    control performance as outcomes of controller synthesis. 
    From a designer perspective, systematic CT-RL synthesis often degenerates to a haphazard trial-and-error search over a high-dimensional and unintuitive hyperparameter space. 

    \item\textbf{CT-RL PE Issues: \cite[Sections IX, X]{BA_Wallace_J_Si_CT_RL_review:2022}.} To achieve PE, CT-RL algorithms almost universally leave the designer only a probing noise $d$ injected at the plant input. However, the natural plant-input disturbance rejection properties which make a controller desirable from a classical perspective render the same controller unfavorable for PE -- revealing a fundamental conflict between classical/RL control principles.
 
\end{enumerate}

\noindent 


%
%


%
%

%
%

\noindent\textbf{Characterization of New Algorithms: Two Novel Constructive RL Design Elements.}
This work proposes 
a numerics-driven, designer-centric framework to improve algorithm learning quality -- the first such approach in CT-RL. 
First, the MI solution realigns the RL excitation framework with classical input/output insights. This creates effective PE while greatly simplifying the process.
%
Second, for systems which exhibit a physically-motivated partition into distinct dynamical loops, our proposed decentralization framework breaks the optimal control problem 
into lower-dimensional subproblems in each of the loops, thereby reducing numerical complexity and dimensionality. 

%
%

\noindent\textbf{Theoretical and Real-World Performance Guarantees.}
We leverage Kleinman's well-tested algorithm \cite{Kleinman_AREs:1968} to rigorously prove convergence/stability guarantees of the developed methods. These algorithms output an optimal LQR controller $K^{*}$ associated with the linearization of the nonlinear dynamics under consideration.
%
We furthermore corroborate these theoretical results by demonstrating practical performance guarantees on a challenging real-world unstable, nonminimum phase HSV example \cite{Wang_Stengel_HSV_tracking_robust:2000,Marrison_Stengel_HSV_tracking_robust:1998,Shaughnessy_Pinckney_McMinn_Cruz_Kelley_HSV_modeling_NASA_Langley:1990} with significant model uncertainty.

%
%

%


%
%

\noindent\textbf{Broad Applicability of Proposed Frameworks.}
We keep the MI framework formulation general, illustrating that this idea may be readily applied to most existing ADP-based RL control methods for PE/conditioning improvements. Meanwhile, a variety of compelling real-world applications admit natural decentralizing dynamical partitions. 
In particular, the longitudinal dynamics of the HSV model studied here naturally separate into a translational/velocity loop and a rotational/flightpath angle loop \cite{Bolender_Doman_HSV_modeling_FPA:2006,Bolender_Doman_HSV_modeling:2005,Bolender_Doman_HSV_modeling:2007}. This translational/rotational decentralization has demonstrated great success in classically-based HSV control methods \cite{Dickeson_AA_Rodriguez_Sridharan_etal_HSV_decentralized_control:2009,Dickeson_AA_Rodriguez_Sridharan_etal_HSV:2009,Dickeson_PhD_thesis_ASU:2012,Marrison_Stengel_HSV_tracking_robust:1998,Wang_Stengel_HSV_tracking_robust:2000,Parker_Serrani_Yurkovich_Bolender_Doman_HSV_tracking_feedback_lin:2006} and applies to general aviation systems, a wide-reaching field of control applications \cite{Stengel_flight_dynamics:book:2022}.
In robotics applications, the Euler-Lagrange equations partition states along the system degrees of freedom \cite{Craig_robotics:book:2005}. Ground robot dynamics, for instance, decompose into a translational speed loop and a rotational steering loop \cite{Dhaouadi_Abu_Hatab_DDMR_modeling:2013,K_Mondal_PhD_thesis:2021}. 
Helicopter dynamics partition along each of the vehicle's three translational and three rotational degrees of freedom \cite{Enns_J_Si_Apache_NDP:2002,Enns_Si_helicopter:2003,Enns_J_Si_helicopter_rotor_failure:2003}, as do Quadcopter/UAV dynamics \cite{Wang_Man_Cao_Zheng_Zhao_quadcopter:2016}.

%
%


\noindent\textbf{Contributions.}
1) We introduce a novel RL design paradigm with two important decentralization and multi-injection constructive elements to systematically improve PE/numerics. 
2) We develop a novel suite of (decentralized) EIRL algorithms combining these elements, and we systematically demonstrate how application of each element successively improves conditioning performance relative to existing ADP methods, 
 making these algorithms capable of realistic and complex controller synthesis.
3) Leveraging classical control insights, we prove rigorous theoretical convergence and closed-loop stability guarantees of the new algorithms.

%
%

\noindent\textbf{Organization.} 
We first define the basic features of each of the methods in Section \ref{sec:problem_formulation}. We then develop the dEIRL algorithm in section \ref{sec:algs_training}, proving convergence and stability in Section \ref{sec:theoretical_results}. Section \ref{sec:ES} provides comprehensive quantitative performance evaluations. Finally, we conclude this study with a discussion in Section \ref{sec:conclusion}.



%
%


%
%

\section{Problem Formulation}\label{sec:problem_formulation}

\noindent\textbf{System.} We consider the continuous-time nonlinear time-invariant affine system
\begin{align}
    \dot{x}
    =
    f(x) + g(x) u,
    \label{eq:sys_nonlin}
\end{align}
\noindent where $x \in \mathbb{R}^{n}$ is the state vector, $u \in \mathbb{R}^{m}$ is the control vector, $f : \mathbb{R}^n \rightarrow \mathbb{R}^n$, and $g : \mathbb{R}^n \rightarrow \mathbb{R}^{n \times m}$. We assume that $f$ and $g$ are Lipschitz on a compact set $\Omega \subset \mathbb{R}^n$ containing the origin $x = 0$ in its interior, and that $f(0) = 0$. In the case of decentralization, we further assume the system (\ref{eq:sys_nonlin}) affords a physically-motivated dynamical partition of the form
\begin{align}
	\left[
	\begin{array}{c}
		\dot{x}_{1}
		\\
		\dot{x}_{2}
	\end{array}
	\right]
	=
	\left[
	\begin{array}{c}
		f_{1}(x)
		\\
		f_{2}(x)
	\end{array}
	\right]	
	+
	\left[
	\begin{array}{cc}
		g_{11}(x) & g_{12}(x)
		\\
		g_{21}(x) & g_{22}(x)
	\end{array}
	\right]	
	\left[
	\begin{array}{c}
		u_{1}
		\\
		u_{2}
	\end{array}
	\right],		
	\label{eq:sys_nonlin_2x2}
\end{align}
\noindent where $x_{j} \in \mathbb{R}^{n_{j}}$, $u_{j} \in \mathbb{R}^{m_{j}}$ $(j = 1, 2 \triangleq N)$ with $n_{1} + n_{2} = n$, $m_{1} + m_{2} = m$. For convenience, we define $g_{j} : \mathbb{R}^{n} \rightarrow \mathbb{R}^{n_{j} \times m}$,  $g_{j}(x) = \left[ \begin{array}{cc} g_{j1}(x) & g_{j2}(x) \end{array} \right] $.
%
%
\begin{remark}[dEIRL Extensible to $N > 2$ Loops]\label{rk:decentralization_more_2_loops}
\noindent We carry on the discussion with $N = 2$ loops for simplicity of presentation, but the structure and theoretical results generalize to $N > 2$ loops. 
\end{remark}

%
%

\begin{table}[h]
	\caption{Data \& Dynamical Information Required}
	\begin{minipage}{0.5\textwidth}	
	\centering
	\begin{tabular}{|c||c|c||c|c|}
	        \hline
	        \multirow{2}{*}{System Type}  & \multicolumn{2}{c||}{EIRL} & \multicolumn{2}{c|}{dEIRL Loop $j$}
	        \\
	        \hhline{|~||-|-||-|-|}
	         &  Data & Dyn\footnote{For definitions: $w$ (\ref{eq:nonlin_sys_irl_rewritten}), $g$ (\ref{eq:sys_nonlin}), $B$ (\ref{eq:sys_lin}).} & Data & Dyn\footnote{For definitions: $w_{j}$ (\ref{eq:nonlin_sys_dirl_rewritten}), $g_{j}$, $g_{jj}$ (\ref{eq:sys_nonlin_2x2}), $A_{jk}$, $B_{j}$, $B_{jj}$ (\ref{eq:sys_lin_2x2}).}
	        \\
	        \hline 
	        \hline
	        Nonlin, Coupled & $(x, u)$ & $w$, $g$ & $(x, u)$ & $w_{j}$, $g_{j}$
	        \\
	        \hline
			Lin, Coupled & $(x, u)$ & $B$ & $(x, u)$ & $A_{jk}$ ($k \neq j$), $B_{j}$
	        \\
	        \hline	 
			Nonlin, Decoup & $(x, u)$ & $w$, $g$ & $(x_{j}, u_{j})$ & $w_{j}$, $g_{jj}$
	        \\
	        \hline	
			Lin, Decoup & $(x, u)$ & $B$ & $(x_{j}, u_{j})$ & $B_{jj}$
	        \\
	        \hline	         
	\end{tabular}
	\vspace{-0.3cm}	
	\end{minipage}	
	\label{tb:data_dyn_reqd}
\end{table}

%
%

\begin{remark}[Dynamical Information of Physical Process Required]
%
%
Table \ref{tb:data_dyn_reqd} summarizes the state-action data and dynamical information required to run EIRL and decentralized EIRL (dEIRL) in loop $1 \leq j \leq N$. In general, both algorithm types require knowledge of the system input dynamics $g$, which is a usual feature of IRL \cite{Vrabie_Lewis_IRL:2009}. They also require knowledge of the \emph{drift difference} $w$ (\ref{eq:nonlin_sys_irl_rewritten}), $w_{j}$ (\ref{eq:nonlin_sys_dirl_rewritten}), respectively. 
In the case the system is linear, EIRL does not require knowledge of the drift dynamics $A$. Meanwhile for dEIRL,
in the case that the system is linear of the form (\ref{eq:sys_lin_2x2}), then $w_{j} = \sum_{k \neq j} A_{jk} x_{k}$. Knowledge of the dominant diagonal drift dynamics term $A_{jj}$ is no longer required; rather, the designer only requires knowledge of the off-diagonal dynamics $A_{jk}$.
In the case that the system (\ref{eq:sys_nonlin_2x2}) is decoupled, $f_{j}$ and hence $w_{j}$ (\ref{eq:nonlin_sys_dirl_rewritten}) are functions of $x_{j}$ only. Furthermore, in this case $g_{j}(x) u = g_{jj}(x_{j}) u_{j}$. Altogether, this says that only collection of state-action data $(x_{j}, u_{j})$ is required to run dEIRL in loop $j$, rather than the entire dataset $(x, u)$. Thus, dynamical decoupling implies algebraic decoupling in the dEIRL algorithm.
Finally, in the dEIRL algorithm, we solve the regression (\ref{eq:DIRL_nonlin_lsq}) by using a nominal model $\tilde{f}_{j}$, $\tilde{w}_{j}$, $\tilde{A}_{jj}$ instead of the actual model $f_{j}$, $w_{j}$, $A_{jj}$, which may contain unknown modeling parameters and is generally not available. We numerically demonstrate this capability in Section \ref{ssec:ES_CL_error}.
\end{remark}

%
%

\noindent\textbf{LQR Problem (Background).}
In the LQR problem, we consider the continuous-time linear time-invariant system
\begin{align}
	\dot{x}
	=
	A x + B u,
	\label{eq:sys_lin}
\end{align}
\noindent where $x \in \mathbb{R}^{n}$, $u \in \mathbb{R}^{m}$ are the state and control vectors, respectively, $A \in \mathbb{R}^{n \times n}$, and $B \in \mathbb{R}^{n \times m}$. We assume that $(A, B)$ is stabilizable \cite{AAR_multivariable:book}, and we denote $(A, B)$ as the linearization of $(f, g)$ (\ref{eq:sys_nonlin}). In the case of decentralization, we partition $(A, B)$ with the appropriate dimensions
\begin{align}
	\left[
	\begin{array}{c}
		\dot{x}_{1}
		\\
		\dot{x}_{2}
	\end{array}
	\right]
	=
	\left[
	\begin{array}{cc}
		A_{11} & A_{12}
		\\
		A_{21} & A_{22}
	\end{array}
	\right]	
	\left[
	\begin{array}{c}
		x_{1}
		\\
		x_{2}
	\end{array}
	\right]
	+
	\left[
	\begin{array}{cc}
		B_{11} & B_{12}
		\\
		B_{21} & B_{22}
	\end{array}
	\right]	
	\left[
	\begin{array}{c}
		u_{1}
		\\
		u_{2}
	\end{array}
	\right],		
	\label{eq:sys_lin_2x2}
\end{align}
and analogously, we define $B_{j} = \left[ \begin{array}{cc} B_{j1} & B_{j2} \end{array} \right] \in \mathbb{R}^{n_{j} \times m}$.
LQR considers the quadratic cost functional
\begin{align}
	J(x_{0})
	=
	\int_{0}^{\infty} (x^{T} Q x + u^{T} R u) \, d \tau,
	\label{eq:LQR_cost_funct}
\end{align}  
\noindent where $Q \in \mathbb{R}^{n \times n}$, $Q = Q^{T} \geq 0$ and $R \in \mathbb{R}^{m \times m}$, $R = R^{T} > 0$ are the state and control penalty matrices, respectively. It is assumed that $(Q^{1/2}, A)$ is detectable \cite{AAR_multivariable:book}. For decentralization, we naturally impose the block-diagonal Q-R cost structure
\begin{align}
	Q 
	=
	\left[
	\begin{array}{cc}
		Q_{1} & 0
		\\
		0 & Q_{2}
	\end{array}
	\right]	,
	\quad
	R 
	=
	\left[
	\begin{array}{cc}
		R_{1} & 0
		\\
		0 & R_{2}
	\end{array}
	\right],	
    \label{eq:LQR_penalties_2x2}
\end{align}
\noindent where $Q_{j} \in \mathbb{R}^{n_{j} \times n_{j}}$, $Q_{j} = Q_{j}^{T} \geq 0$, and $R_{j} \in \mathbb{R}^{m_{j} \times m_{j}}$, $R_{j} = R_{j}^{T} > 0$ $(j = 1, 2)$.
%
%
%
%
%
Under the above dynamical assumptions, the LQR optimal control $u^{*}$ associated with the quadruple $(A, B, Q, R)$ exists, is unique, and assumes the form of a full-state feedback control law \cite{AAR_multivariable:book}
\begin{align}
	u^{*}
	=
	- K^{*} x,
    \label{eq:Kstar_LQR}
\end{align}

\noindent where $K^{*} = R^{-1} B^{T} P^{*}$, and $P^{*} \in \mathbb{R}^{n \times n}$, $P^{*} = {P^{*}}^{T} > 0$ is the unique positive definite solution to the control algebraic Riccati equation (CARE)
\begin{align}
	A^{T} P^{*} + P^{*} A - P^{*} B R^{-1} B^{T} P^{*} + Q
	=
	0.
	\label{eq:CARE}
\end{align}

%
%

\noindent\textbf{Kleinman's Algorithm \cite{Kleinman_AREs:1968} (Background).}
 Suppose that $K_{0} \in \mathbb{R}^{m \times n}$ is such that $A - B K_{0}$ is Hurwitz. For iteration $i$ $(i = 0, 1, \dots)$, let $P_{i} \in \mathbb{R}^{n \times n}$, $P_{i} = P_{i}^{T} > 0$ be the symmetric positive definite solution of the algebraic Lyapunov equation (ALE)
\begin{align}
	(A - B K_{i})^{T} P_{i} + P_{i} (A - B K_{i}) + K_{i}^{T} R K_{i} + Q
	=
	0.
	\label{eq:Kleinman_LE}
\end{align}
\noindent Having solved the ALE $P_{i}$ (\ref{eq:Kleinman_LE}), the controller $K_{i+1} \in \mathbb{R}^{m \times n}$ is updated recursively as
\begin{align}
	K_{i+1}
	=
	R^{-1} B^{T} P_{i}.
	\label{eq:Kleinman_controller_update}
\end{align}

%
%

\noindent\textbf{Relevant Operators.}
%

%
%

\begin{definition}
Let $\underline{n} = \tensor*[_{n}]{P}{_{2}} = \frac{n(n+1)}{2}$. Define the maps $v : \mathbb{R}^{n \times n} \rightarrow \mathbb{R}^{\underline{n}}$, and $\mathcal{B} : \mathbb{R}^{n} \times \mathbb{R}^{n} \rightarrow \mathbb{R}^{\underline{n}}$ by
\begin{align}
	&
	v(P)
	=
	\big[
	p_{11}, \, 2 p_{12}, \dots, \, 2 p_{1n}, 
	\nonumber
	\\
	&\qquad\qquad\qquad
	p_{22}, \, 2 p_{23}, \dots, \, 2 p_{n-1,n}, \, p_{nn}
	\big]^{T},
	\label{eq:vP_def}	
	\\
	&
	\mathcal{B}(x,y)
	=
	\frac{1}{2}
	\big[
	2 x_{1} y_{1}, \, x_{1} y_{2} + x_{2} y_{1}, \dots, \, x_{1} y_{n} + x_{n} y_{1}, 
	\nonumber
	\\
	&
	\, 2 x_{2} y_{2}, \, x_{2} y_{3} + x_{3} y_{2}, \dots, \, x_{n-1} y_{n} + x_{n} y_{n-1}, \, 2 x_{n} y_{n}
	\big]^{T}.
	\label{eq:Bxy_def}		
\end{align}

\noindent Define $W \in \mathbb{R}^{\underline{n} \times n^{2}}$ as the matrix satisfying the identity
\begin{align}
	\mathcal{B}(x, y)
	=
	W (x \otimes y),
	\quad
	\forall \; x, y \in \mathbb{R}^{n},
	\label{eq:Bxy_M_id}	
\end{align}

\noindent where $\otimes$ denotes the Kronecker product \cite{Brewer_Kronecker_products:1978}. For $l \in \mathbb{N}$ and a strictly increasing sequence $\{t_{k}\}_{k=0}^{l}$, whenever $x, y : [t_{0}, t_{l}] \rightarrow \mathbb{R}^{n}$, define the matrix $\delta_{xy} \in \mathbb{R}^{l \times \underline{n}}$ as
\begin{align}
	\delta_{xy}
	&=
	\left[
	\begin{array}{c}
	\mathcal{B}^{T}\big( x(t_{1}) + y(t_{0}), x(t_{1}) - y(t_{0}) \big)
	\\
	\mathcal{B}^{T}\big( x(t_{2}) + y(t_{1}), x(t_{2}) - y(t_{1}) \big)
	\\
	\vdots
	\\
	\mathcal{B}^{T}\big( x(t_{l}) + y(t_{l-1}), x(t_{l}) - y(t_{l-1}) \big)	
	\end{array}
	\right].
\end{align}

\noindent Whenever $x, y$ are square-integrable, define $I_{\mathcal{B}(x,y)} \in \mathbb{R}^{l \times \underline{n}}$ as
\begin{align}
	I_{\mathcal{B}(x,y)}
	&=
	\left[
	\begin{array}{c}
	\int_{t_{0}}^{t_{1}} \mathcal{B}^{T}(x,y) \, d\tau
	\\
	\int_{t_{1}}^{t_{2}} \mathcal{B}^{T}(x,y) \, d\tau
	\\
	\vdots
	\\
	\int_{t_{l-1}}^{t_{l}} \mathcal{B}^{T}(x,y) \, d\tau
	\end{array}
	\right].	
	\label{eq:I_Bxy_def}
\end{align}

\end{definition}

%
%

\begin{proposition}\label{prop:operator_propts}
The operators $v$ (\ref{eq:vP_def}), $\mathcal{B}$ (\ref{eq:Bxy_def}), and matrix $W$ (\ref{eq:Bxy_M_id}) satisfy the following:

\begin{enumerate}[(i)]

	\item $v$ is a linear surjection whose kernel is the subspace of strictly-lower-triangular matrices. Thus, the restriction of $v$ to the symmetric matrices is a linear isomorphism. 
	
	\item $\mathcal{B}$ is a symmetric bilinear form.
	
	\item Whenever $P \in \mathbb{R}^{n \times n}$, $P = P^{T}$, the following identity holds
	\begin{align}
		\mathcal{B}^{T}(x, y) v(P) = x^{T} P y,
		\quad 
		\forall \; x, y \in \mathbb{R}^{n}.
		\label{eq:Bxy_P_id}
	\end{align}
	
	\item $\norm{W} = 1$, and the rows of $W$ are nonzero and pairwise orthogonal. In particular, $W$ has a right inverse which we shall denote $W_{r}^{-1} \in \mathbb{R}^{n^{2} \times \underline{n}}$ satisfying the identity
	\begin{align}
		x \otimes x
		=
		W_{r}^{-1} \mathcal{B}(x,x),
		\quad 
		\forall \; x \in \mathbb{R}^{n}.
		\label{eq:Bxx_Mr_id}
	\end{align}
	
\end{enumerate}

\end{proposition}



%
%


%
%

%
%
\renewcommand{\relpath}{figures/algs_training/}

\section{Algorithms \& Training}\label{sec:algs_training}

In the classical LQR problem, Kleinman's algorithm \cite{Kleinman_AREs:1968} reduces the CARE, a quadratic matrix equation, to an iterative sequence of algebraic Lyapunov equations (ALEs), linear matrix equations. Inspired by Kleinman's approach, EIRL and dEIRL use state-action data generated by the \emph{nonlinear} system (\ref{eq:sys_nonlin}) to iteratively solve the CARE associated with its linearization via a sequence of simpler linear regression problems. dEIRL reduces the dimension of the regressions by taking advantage of the decentralized dynamical structure (\ref{eq:sys_nonlin_2x2}). 

We first note that EIRL and dEIRL can each be implemented with single-injection (SI) and multi-injection (MI) modes. As such, we have a suite of four CT-RL algorithms in this paper. 
Figure \ref{fig:classical_fb} shows a standard negative feedback structure, consisting of a controller $K$ and plant $P$ (each of which may be linear or nonlinear). For SI, a probing noise $d$ is injected at the plant input (i.e., at the same location as the input disturbance $d_{i}$ in Figure \ref{fig:classical_fb}). This is the traditional injection point for CT-RL algorithms \cite{BA_Wallace_J_Si_CT_RL_review:2022}. In the case of MI, a reference command $r$, selected by the designer to achieve PE, is injected in addition to the usual probing noise $d$. We will discuss these two modes further in Sections \ref{ssec:multi_injection_conflict} and \ref{ssec:multi_injection}.

%
%

\setlength{\unitlength}{.09in}
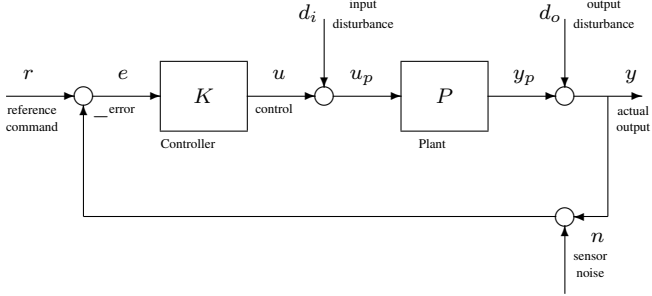
\begin{figure}[h]
	\begin{center}
    \hspace{0.15in}
		\begin{picture}(31,15)
		\footnotesize
		%
		%
		
		\put(-3,7){$r$}                                             
		\put(-4,4){\shortstack{\tiny{reference}\\ \tiny{command}}}  
		\put(-4,6){\vector(1,0){4}}                                 
		
		%
		%
		\put(1,6){\vector(1,0){4}}                    
		\put(2.5,7){$e$}                              
		\put(2,5){\shortstack{\tiny{error}}}          

		\put(5,4){\framebox(5,4){$K$}}                
		\put(5,3){\tiny{Controller}}                  
		\put(10,6){\vector(1,0){4}}                   
		\put(11.5,7){$u$}                             
		\put(10.5,5){\shortstack{\tiny{control}}}     

		%
		%
		\put(14.5,6){\circle{1}}                      
		\put(14.5,10.5){\vector(0,-1){4}}             
		\put(13,10.5){$d_i$}                          
		\put(15,10){\shortstack{\tiny{input}\\\tiny{disturbance}}}  

		%
		%
		\put(15,6){\vector(1,0){4}}                   
		\put(16,7){$u_p$}                             
		\put(19,4){\framebox(5,4){$P$}}               
		\put(20,3){\tiny{{Plant}}}                
		\put(24,6){\vector(1,0){4}}                   
		\put(25.5,7){$y_p$}                           
		
		%
		%
		\put(28.5,6){\circle{1}}                      
		\put(28.5,10.5){\vector(0,-1){4}}             
		\put(27,10.5){$d_o$}                          
		\put(29,10){\shortstack{\tiny{output}\\\tiny{disturbance}}} 
		
		%
		%
		\put(29,6){\vector(1,0){4}}                                 
		\put(32,7){$y$}                                             
		\put(31.5,4){\shortstack{\tiny{actual}\\\tiny{output}}}       
		
		%
		%
		\put(31,6){\line(0,-1){7}}                    
		\put(31,-1){\vector(-1,0){2}}                 
		\put(28,-1){\line(-1,0){27.5}}                
		\put(0.5,-1){\vector(0,1){6.5}}               
		\put(1,4.5){$-$}                              
		\put(0.5,6){\circle{1}}                       
		
		%
		%
		\put(28.5,-1){\circle{1}}                     
		\put(28.5,-5.5){\vector(0,1){4}}              
		\put(30,-2.5){$n$}                            
		\put(29,-4.5){\shortstack{\tiny{sensor}\\\tiny{noise}}}     
		
		%
		%

		\end{picture}
		\vspace*{0.4in}
		\caption{Standard negative feedback structure.}
		\label{fig:classical_fb}
	\end{center}
\end{figure}

In the following sections, suppose that $K_{0} \in \mathbb{R}^{m \times n}$ is such that $A - B K_{0}$ is Hurwitz, and analogously that $K_{0,j} \in \mathbb{R}^{m_{j} \times n_{j}}$ is such that $A_{jj} - B_{jj} K_{0,j}$ is Hurwitz in loop $j$ ($j = 1, \dots, N$). Let an iteration $i \geq 0$ be given.

%
%

\subsection{Single-Injection Excitable Integral Reinforcement Learning (SI-EIRL)}\label{ssec:irl_forumlation}

Suppose the control $u = - K_{0} x + d$ is applied to the system (\ref{eq:sys_nonlin}), where $d : \mathbb{R}_{+} \rightarrow \mathbb{R}^{m}$ is an exploration noise. Manipulating (\ref{eq:sys_nonlin}), we have
\begin{align}
	\dot{x}
	&=
	A_{i} x + B K_{i} x + g(x) u + w,
	\nonumber
	\\
	w
	&\triangleq
	f(x) - A x,
	\qquad
	A_{i}
	\triangleq
	A - B K_{i}.
	\label{eq:nonlin_sys_irl_rewritten}
\end{align}

\noindent Let $t_{0} < t_{1}$. Along the solutions of (\ref{eq:nonlin_sys_irl_rewritten}), we have
\begin{align}
	&
	x^{T}(t_{1}) P_{i} x(t_{1}) - x^{T}(t_{0}) P_{i} x(t_{0})
	=
	\int_{t_{0}}^{t_{1}} \bigg[ x^{T} \big( A_{i}^{T} P_{i} + P_{i} A_{i} \big) x 
	\nonumber
	\\
	&\qquad	
	+ 2 (B K_{i} x + g(x) u + w)^{T} P_{i} x \bigg] \, d\tau.
	\label{eq:IRL_nonlin_single_sample_id}
\end{align}

\noindent Applying (\ref{eq:Bxy_P_id}) and rearranging terms, (\ref{eq:IRL_nonlin_single_sample_id}) becomes
\begin{align}
	&
	\bigg[ \mathcal{B}\big( x(t_{1}) + x(t_{0}), x(t_{1}) - x(t_{0}) \big) 
	\nonumber
	\\
	&\quad
	- 2 \int_{t_{0}}^{t_{1}} \mathcal{B}\big( B K_{i} x + g(x) u + w, x \big) \, d\tau  \bigg]^{T} v(P_{i})
	\nonumber
	\\
	&\qquad\qquad=
	\left[ \int_{t_{0}}^{t_{1}} \mathcal{B}\big( x, x \big) \, d\tau \right]^{T} v \big( A_{i}^{T} P_{i} + P_{i} A_{i} \big) 
	\label{eq:IRL_nonlin_single_sample_before_LE}
	\\
	&\qquad\qquad=	
	-\left[ \int_{t_{0}}^{t_{1}} \mathcal{B}\big( x, x \big) \, d\tau \right]^{T} v \big( Q + K_{i}^{T} R K_{i} \big),
	\label{eq:IRL_nonlin_single_sample}
\end{align}

\noindent where the last equality (\ref{eq:IRL_nonlin_single_sample}) follows from the fact that $P_{i} = P_{i}^{T} > 0$ satisfies the ALE (\ref{eq:Kleinman_LE}). For $l \in \mathbb{N}$ and a strictly increasing sequence $\{t_{k}\}_{k=0}^{l}$, applying (\ref{eq:Bxy_M_id}), (\ref{eq:Bxx_Mr_id}), (\ref{eq:IRL_nonlin_single_sample}) and manipulating further, we arrive at the least-squares regression
\begin{align}
	\Theta_{i} v(P_{i}) 
	&=
	\Xi_{i},
	\label{eq:IRL_nonlin_lsq}
\end{align}

\noindent where $\Theta_{i} \in \mathbb{R}^{l \times \underline{n}}$, $\Xi_{i} \in \mathbb{R}^{l}$ are given by
\begin{align}
	\Theta_{i}
	&=
	\delta_{xx} - 2 \big[  I_{\mathcal{B}(x, x)} W_{i}^{T}  + I_{\mathcal{B}(x, gu)} + I_{\mathcal{B}(x, w)} \big],
    \nonumber
    \\
    W_{i}
	&\triangleq	
	W \left( I_{n} \otimes B K_{i} \right) W_{r}^{-1},  
	\label{eq:IRL_nonlin_lsq_A}
	\\
	\Xi_{i}
	&=
	-I_{\mathcal{B}(x, x)} v (Q_{i}), 
	\qquad
	Q_{i} 
	\triangleq
	Q + K_{i}^{T} R K_{i}.
	\label{eq:IRL_nonlin_lsq_b}
\end{align}

Having performed the regression $v(P_{i})$ (\ref{eq:IRL_nonlin_lsq}), we update the controller $K_{i + 1}$ with (\ref{eq:Kleinman_controller_update}).

%
%

\begin{remark}[EIRL Algorithm vs. Original IRL Formulation \cite{Vrabie_Lewis_IRL:2009}: Probing Noise, Data Reuse, Hyperparameters]\label{rk:new_IRL_vs_orig}
Crucially, EIRL accommodates probing noise injection, which IRL \cite{Vrabie_Lewis_IRL:2009} does not allow for. Lack of probing noise proves a practical design hindrance, as it renders proper system excitation nearly impossible \cite{BA_Wallace_J_Si_CT_RL_review:2022}. 
Furthermore, the algebra derived for the term $I_{\mathcal{B}(x, B K_{i} x)} = I_{\mathcal{B}(x, x)} W_{i}^{T}$ (\ref{eq:IRL_nonlin_lsq_A}) allows the EIRL algorithm to reuse the same state trajectory data collected under the single stabilizing controller $K_{0}$ for generation of the sequence $\{K_{i}\}_{i=1}^{\infty}$. This is in contrast to the original IRL formulation \cite{Vrabie_Lewis_IRL:2009}, which for iteration $i$ requires state-action data to be simulated under the stabilizing controller $K_{i}$ before updating to $K_{i+1}$, thus making data reuse impossible.
%
The observed severe numerical sensitivity of existing CT-RL algorithms to basis function selection is another significant design challenge \cite{BA_Wallace_J_Si_CT_RL_review:2022}. EIRL removes this complexity by fixing the basis as the monomials of total degree two $\mathcal{B}(x,x) \in \mathbb{R}^{\underline{n}}$, the basis associated with the LQR problem.

\end{remark}

%
%

\begin{remark}[EIRL Algorithm vs. Subsequent IRL Formulation \cite{Jiang_ZP_Jiang_LQR_IRL:2012}: Accommodating Nonlinear Systems, Dimensionality Reductions]\label{rk:new_IRL_vs_old}
The EIRL formulation is inspired by \cite{Jiang_ZP_Jiang_LQR_IRL:2012}, but EIRL offers various practical improvements. Firstly, EIRL accommodates nonlinear systems, while \cite{Jiang_ZP_Jiang_LQR_IRL:2012} applies to linear systems only. More importantly, comparing the least-squares regression (\ref{eq:IRL_nonlin_lsq}) with \cite[Equation (11)]{Jiang_ZP_Jiang_LQR_IRL:2012}, we see that (\ref{eq:IRL_nonlin_lsq}) is lower-dimensional ($\underline{n}$ here versus $\underline{n} + mn$ in \cite{Jiang_ZP_Jiang_LQR_IRL:2012}). This is because the controller $K_{i+1} \in \mathbb{R}^{m \times n}$ is no longer a part of the regression vector in (\ref{eq:IRL_nonlin_lsq}). As a result, knowledge of the system input dynamics $g$ (and hence $B$) is required in (\ref{eq:IRL_nonlin_lsq}).
In light of observed numerical and dimensional scalability issues associated with existing CT-RL algorithms \cite{BA_Wallace_J_Si_CT_RL_review:2022}, the tradeoff for reduced dimensionality in exchange for system knowledge is a deliberate one. 

\end{remark}

%
%

\subsection{Single-Injection Decentralized Excitable Integral Reinforcement Learning (SI-dEIRL)}\label{ssec:dirl_forumlation}

In this section, we illustrate how EIRL may be readily generalized to a decentralized system of the form (\ref{eq:sys_nonlin_2x2}).
Let any loop $1 \leq j \leq N$ be given. We may then apply Kleinman's algorithm (Section \ref{sec:problem_formulation}), yielding sequences $\{P_{i,j}\}_{i=0}^{\infty}$ in $\mathbb{R}^{n_{j} \times n_{j}}$ and $\{K_{i,j}\}_{i=1}^{\infty}$ in $\mathbb{R}^{m_{j} \times n_{j}}$ from the ALE 
\begin{align}
	(A_{jj} - B_{jj} K_{i,j})^{T} P_{i,j} 
	&+ 
	P_{i,j} (A_{jj} - B_{jj} K_{i,j}) 
	\nonumber
	\\
	&\qquad
	+ K_{i,j}^{T} R_{j} K_{i,j} + Q_{j}
	=
	0.
	\label{eq:Kleinman_LE_decoupled}
\end{align}

\noindent Suppose the control $u_{j} = - K_{0,j} x_{j} + d_{j}$ is applied to the system (\ref{eq:sys_nonlin_2x2}), where $d_{j} : \mathbb{R}_{+} \rightarrow \mathbb{R}^{m_{j}}$ is an exploration noise. Manipulating (\ref{eq:sys_nonlin_2x2}), we have
\begin{align}
	\dot{x}_{j}
	&=
	A_{i,j} x_{j} + B_{jj} K_{i,j} x_{j} + g_{j}(x) u + w_{j},
	\nonumber
	\\
	w_{j}
	&\triangleq
	f_{j}(x) - A_{jj} x_{j},
	\qquad
	A_{i,j}
	\triangleq
	A_{jj} - B_{jj} K_{i,j}.
	\label{eq:nonlin_sys_dirl_rewritten}
\end{align}

\noindent Given $l_{j} \in \mathbb{N}$ and a strictly increasing sequence $\{t_{k,j}\}_{k=0}^{l_{j}}$, following a derivation entirely analogous to that presented in Section \ref{ssec:irl_forumlation}, we arrive at the least-squares regression
\begin{align}
	\Theta_{i,j} v(P_{i,j}) 
	&=
	\Xi_{i,j},
	\label{eq:DIRL_nonlin_lsq}
\end{align}

\noindent where $\Theta_{i,j} \in \mathbb{R}^{l_{j} \times \underline{n}_{j}}$, $\Xi_{i,j} \in \mathbb{R}^{l_{j}}$ are given by
\begin{align}
	\Theta_{i,j}
	&=
	\delta_{x_{j}x_{j}} - 2 \big[  I_{\mathcal{B}(x_{j}, x_{j})} W_{i,j}^{T}  + I_{\mathcal{B}(x_{j}, g_{j}u)} + I_{\mathcal{B}(x_{j}, w_{j})} \big],
    \nonumber
    \\
	W_{i,j}
	&\triangleq	
	W \left( I_{n_{j}} \otimes B_{jj} K_{i,j} \right) W_{r}^{-1}, \label{eq:DIRL_nonlin_lsq_A}
	\\
	\Xi_{i,j}
	&=
	-I_{\mathcal{B}(x_{j}, x_{j})} v (Q_{i,j}), 
	\;\;\;
    Q_{i,j} 
	\triangleq
	Q_{j} + K_{i,j}^{T} R_{j} K_{i,j}.
	\label{eq:DIRL_nonlin_lsq_b}
\end{align}

Having performed the regression $v(P_{i,j})$ (\ref{eq:DIRL_nonlin_lsq}), we update the controller analogously to (\ref{eq:Kleinman_controller_update}):
\begin{align}
    K_{i+1,j} 
    = 
    R_{j}^{-1} B_{jj}^{T} P_{i,j}.
    \label{eq:DIRL_controller_update}
\end{align}

%
%

\subsection{Probing Noise Injection: Insights on a Fundamental Conflict between RL \& Classical Control Principles}\label{ssec:multi_injection_conflict}

PE requirements are often invoked in proofs of CT-RL algorithm convergence \cite{Vrabie_Lewis_IRL:2009,Vamvoudakis_Lewis_SPI:2010,Jiang_ZP_Jiang_Robust_ADP:2014,Bian_ZP_Jiang_ADP_VI:2021}, and to achieve PE it has long been standard practice to apply a probing noise $d$ to the system (\ref{eq:sys_nonlin}). Unfortunately, the empirical analysis in \cite{BA_Wallace_J_Si_CT_RL_review:2022} illustrates that the requirement of PE introduces myriad practical design complications, even for simple academic second-order examples. Severe empirical problems often indicate an underlying violation of principle. 

\begin{figure*}[h]
    \centering
    \subfloat[]{\includegraphics[height=\figsize]{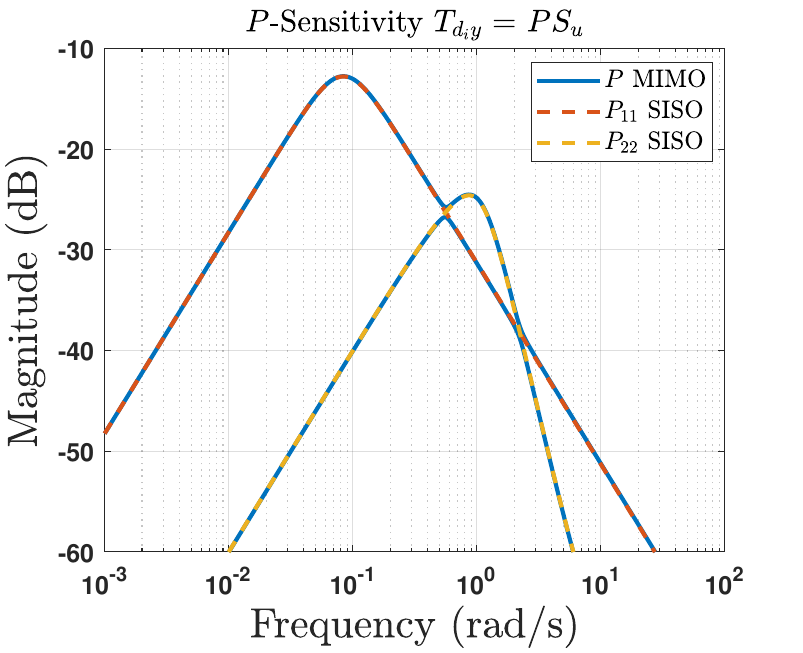}%
    \label{fig:PSu}}
    \hfil
    \subfloat[]{\includegraphics[height=\figsize]{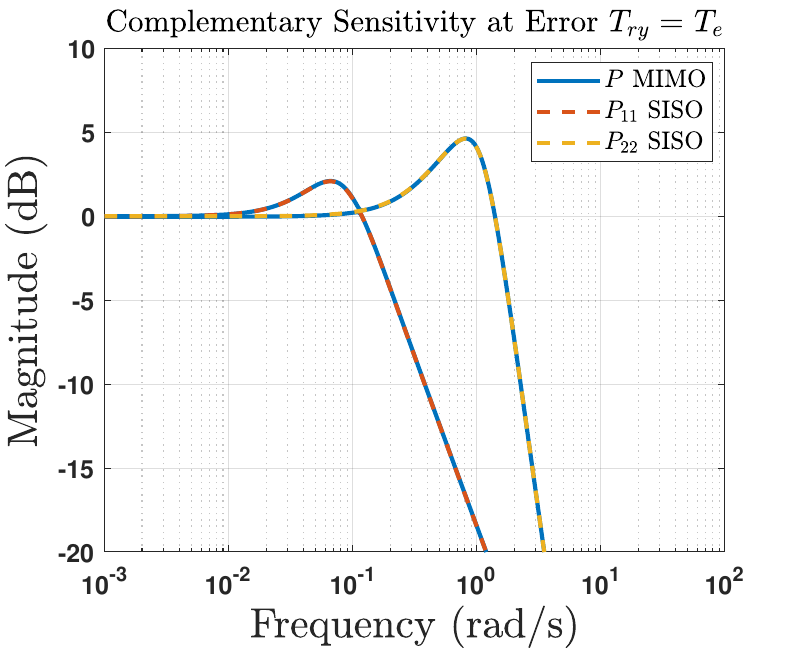}%
    \label{fig:Te}}
    \caption{Closed-loop frequency responses: The probing noise injection issue, visualized. 
    (a): $P$-sensitivity $P S_{u} = T_{d_{i} y}$. 
    (b): Complementary sensitivity at the error $T_{e} = T_{r y}$.}
    \label{fig:PSu_vs_Te}
\end{figure*}

To gain insights on the issue, we first define a few relevant closed-loop maps. Returning to Figure \ref{fig:classical_fb} in the case of a linear plant $P$ and controller $K$, we define the $P$-sensitivity $P S_{u} = T_{d_{i} y}$ as the closed-loop map from plant input disturbance $d_{i}$ to plant output $y$, and we define the complementary sensitivity at the error $T_{e} = T_{r y}$ as the closed-loop map from reference command $r$ to plant output $y$ \cite{AAR_multivariable:book}. To illustrate typical input-output behavior, Figure \ref{fig:PSu_vs_Te} shows these two closed-loop frequency responses for the nominal HSV design 
in Section \ref{sec:ES} -- both the exact MIMO frequency response (blue solid curve) and SISO approximations in each of the respective $j = 2$ loops (dashed curves).

%
%
%

Let us examine the frequency response in loop $j = 2$ (associated with flightpath angle $y_{2} = \gamma$, yellow dashed curve), which will turn out to be the most numerically-troublesome in Section \ref{sec:ES}.
Since probing noise is inserted at the \emph{plant input}, the effective closed-loop map from probing noise $d$ to output $y$ is the $P$-sensitivity $T_{d_{i} y}$. With this in mind, one glance at the yellow dashed SISO $T_{d_{i} y}$ response in Figure \ref{fig:PSu} immediately reveals issues: Any probing noise will be attenuated by at least $-25$ dB (a factor of about 20) as a best-case, and only so near frequencies of $\omega \approx 1$ rad/s. Even more troubling, any probing noise frequency content below $\omega \approx 10^{-1}$ rad/s and above $\omega \approx 2.5$ rad/s will be attenuated by more than $-40$ dB -- a factor of 100. In light of this simple linear classical analysis, it is no wonder that achieving sufficient system excitation via probing noise injection proves to be a significant issue for the nonlinear flightpath angle learning in loop $j = 2$. 

This real-world example illustrates a broader and more fundamental divergence between RL and classical control. From a classical perspective, a control designer is \emph{pleased} by the $P$-sensitivity in Figure \ref{fig:PSu}; indeed, good input disturbance rejection is a pillar of sound control system design. On the other hand, from an RL perspective, a designer is \emph{troubled} by such a $P$-sensitivity response, for a best-case attenuation of $-25$ dB means that sufficient system excitation is likely impossible to achieve 
within reasonable bounds of control effort. 
Hence, we arrive at the unfortunate conclusion that the now-standard RL practice of probing noise injection has pitted learning requirements \emph{diametrically} at odds with classical control principles. 
This motivates MI designs.

%
%

\subsection{Multi-Injection (MI) EIRL \& dEIRL}\label{ssec:multi_injection}

In light of the fundamental conflict illustrated in Section \ref{ssec:multi_injection_conflict}, the question naturally arises: How may RL control algorithms be modified to accommodate an excitation framework which is in alignment with classical control principles? The remedy lies again in application of simple input-output intuitions. Just as input-disturbance rejection is a pillar of good control system design, so too is (low-frequency) reference command following. Graphically representative of this principle is the complementary sensitivity $T_{r y}$ plotted in Figure \ref{fig:Te}, which remains near 0 dB (i.e., unity $y \approx r$) at low frequencies. We hence pose the following \emph{multi-injection} solution: Instead of inserting only a probing noise $d$ at the plant input, we apply the control $u = - K e + d$, where the excitation $K e = K(r - y)$ (Figure \ref{fig:classical_fb}) results from insertion of a designed reference command $r$. 

MI offers multiple benefits: First and foremost, reference command injection allows the designer to modulate system excitation via the complementary sensitivity map $T_{r y}$, which is much favorable in comparison to the $P$-sensitivity $T_{d_{i} y}$ from an input-output perspective (Figure \ref{fig:PSu_vs_Te}). Second, the designer now has multiple excitation ``knobs" (namely, the usual probing noise $d$ in addition to the reference command $r$) to tweak in order to improve data quality, offering greater intuitiveness and flexibility. 

Lastly, by virtue of lumping the additional reference command term $r$ in the control $u$, MI doesn't explicitly alter RL algorithm theoretical formulations and can be readily implemented on nonlinear compensators in the general RL context. Specifically, suppose a CT-RL algorithm has been formulated under the standard probing noise injection excitation framework; i.e., requires application of a control of the form $u = \mu(x) + d$ for some stabilizing policy $\mu$. Since the optimal control problem already requires full-state information, it is not restrictive to designate a subset of the state $x$ as measurement variables $y$ for reference injection. Suppose that $p \leq n$ variables of the state $x$ have been chosen so, and assume after possibly re-indexing that $x$ has the form $x = \left[ y^{T} \; x_{r}^{T} \right]^{T}$, where $x_{r} \in \mathbb{R}^{n-p}$ denotes the rest of the state. If we apply a reference excitation $r$ to the closed-loop system, note that the control
\begin{align}
    u
    &=
    \mu(e, x_{r}) + d
    =
    \mu(y, x_{r}) + \tilde{d},
    \\
    \tilde{d}
    &\triangleq
    d + \left( \mu(e, x_{r}) - \mu(y, x_{r}) \right)
\end{align}
\noindent is also of the form $u = \mu(x) + \tilde{d}$ required for execution (it is the arbitrariness of the probing noise signal $d$ which allows for this sort of bookkeeping). Thus, MI is a viable candidate for improving PE properties of existing RL methods involving the usual probing noise formulation.

In exchange for this greatly-enhanced MI excitation tool, computational burden has increased slightly if the compensator $K$ is dynamic (which is the case, e.g., when integral augmentation is performed). In this case, the excitation $K e$ needs to be simulated dynamically online, in addition to the usual response $K y$. 
However, we make the important notes that in spite of this slight computational cost, MI does not affect dimensionality of the underlying learning problem, nor does it impose additional requirements on system dynamics knowledge.

To conclude this section, we summarize the EIRL and dEIRL execution procedure in Algorithm \ref{alg:EIRL}, both in their SI and MI modes.

%
%

\begin{algorithm}[h]
	\caption{EIRL/dEIRL Algorithm.}\label{alg:EIRL}
	\begin{algorithmic}[1]
		\State \textbf{\ul{Hyperparameters:}} 
        \begin{itemize}
            \item $N \in \mathbb{N}$ loops ($N = 1$: EIRL, $N > 1$: dEIRL). 
        \end{itemize}
        In each loop $j = 1, \dots, N$:
        \begin{itemize}
            \item Cost structure $Q_{j}$, $R_{j}$, stabilizing controller $K_{0,j}$.
            \item Learning params: $T_{s,j}$, $l_{j}$, $i_{j}^{*}$, $d_{j}$. If MI: $r_{j}$.
        \end{itemize}
  
		\State \textbf{\ul{Data Collection:}} For $j = 1, \dots, N$, apply
        \If{SI}
            \State $d_{j}$
        \ElsIf{MI}
            \State $d_{j}$, $r_{j}$ 
        \EndIf
        
        \noindent to system (\ref{eq:sys_nonlin}) under $K_{0,j}$, collecting state-action data $\left\{\big( x_{j}(k T_{s,j}), u_{j}(k T_{s,j}) \big) \right\}_{k=0}^{l_{j}}$.

        \State \textbf{\ul{Learning:}}
        \For{$j = 1:N$}
    		\For{$i = 0:i_{j}^*$} 
    		    \State Perform regression $v(P_{i,j})$ (\ref{eq:DIRL_nonlin_lsq}).
                \State Perform inversion $v(P_{i,j}) \mapsto P_{i,j}$ (Prop. \ref{prop:operator_propts}).
                \State Update controller $K_{i,j} \gets K_{i+1,j}$ (\ref{eq:DIRL_controller_update}).
    		\EndFor
        \EndFor
        
		\State\textbf{\ul{Termination:}} Final controllers $K_{i_{j}^*,j}$ ($j = 1,\dots,N$).
	\end{algorithmic}
\end{algorithm}



%
%


%
%

\section{Theoretical Results}\label{sec:theoretical_results}

In this section, we prove key convergence and stability guarantees of the presented methodologies. Throughout, we assume that the baseline dynamical assumptions outlined in Section \ref{sec:problem_formulation} hold. We begin with Kleinman's algorithm.

%
%

\begin{theorem}[Stability, Convergence of Kleinman's Algorithm \cite{Kleinman_AREs:1968}]\label{thm:Kleinman_AREs}
Let the assumptions of Section \ref{sec:problem_formulation} hold. Then we have the following:

\begin{enumerate}[(i)]

	\item $A - B K_{i}$ is Hurwitz for all $i \geq 0$.
	
	\item $P^{*} \leq P_{i+1} \leq P_{i}$ for all $i \geq 0$.
	
	\item $\lim\limits_{i \rightarrow \infty} K_{i} = K^{*}$, $\lim\limits_{i \rightarrow \infty} P_{i} = P^{*}$.
	
\end{enumerate}

\end{theorem}

%
%

We now move on to EIRL. Before proceeding to the main theoretical results, we require the following two lemmas:

\begin{lemma}\label{lem:IRL_nonlin_Pi_lsq}
Suppose that the controller $K_{i} \in \mathbb{R}^{m \times n}$ is stabilizing, and that the matrix $\Theta_{i} \in \mathbb{R}^{l \times \underline{n}}$ (\ref{eq:IRL_nonlin_lsq_A}) has full column rank. Then $P_{i} \in \mathbb{R}^{n \times n}$, $P_{i} = P_{i}^{T} > 0$ is the unique positive definite solution to the ALE (\ref{eq:Kleinman_LE}) if and only if $P_{i}$ satisfies the least-squares regression (\ref{eq:IRL_nonlin_lsq}) at equality. In particular, the least-squares solution of the EIRL algorithm (\ref{eq:IRL_nonlin_lsq}) yields the solution of the associated ALE (\ref{eq:Kleinman_LE}).

\end{lemma}

%
%
\textit{Proof:} 
The forward direction was proved in (\ref{eq:IRL_nonlin_single_sample_id})--(\ref{eq:IRL_nonlin_single_sample}). Conversely, suppose $v(P) \in \mathbb{R}^{\underline{n}}$ minimizes the least-squares regression (\ref{eq:IRL_nonlin_lsq}). Since $\Theta_{i}$ has full column rank, the solution $v(P) \in \mathbb{R}^{\underline{n}}$ is unique. Moreover, letting $P_{i} = P_{i}^{T} > 0$ be the unique positive definite solution to the ALE (\ref{eq:Kleinman_LE}), we have seen that $v(P_{i}) \in \mathbb{R}^{\underline{n}}$ satisfies (\ref{eq:IRL_nonlin_lsq}) at equality. Thus, $v(P) = v(P_{i})$. Since $v$ restricted to the symmetric matrices is a bijection (Proposition \ref{prop:operator_propts}), this implies $P = P_{i}$ is the solution to the ALE (\ref{eq:Kleinman_LE}).
$\hfill\blacksquare$

%
%

\begin{lemma}\label{lem:IRL_nonlin_Thetai_full_rank}
Suppose that $l \in \mathbb{N}$ and the sample instants $\{t_{k}\}_{k=0}^{l}$ are chosen such that the matrix $I_{\mathcal{B}(x,x)}$ (\ref{eq:I_Bxy_def}) has full column rank $\underline{n}$. If $K_{i}$ is stabilizing, then the matrix $\Theta_{i} \in \mathbb{R}^{l \times \underline{n}}$ (\ref{eq:IRL_nonlin_lsq}) has full column rank $\underline{n}$.

\end{lemma}

%
%
\textit{Proof:} 
Follows similarly from the proof of \cite[Lemma 6]{Jiang_ZP_Jiang_LQR_IRL:2012}. Suppose $v(P) \in \mathbb{R}^{\underline{n}}$ is such that $\Theta_{i} v(P) = 0$. Then the identities (\ref{eq:Bxy_P_id}) and (\ref{eq:IRL_nonlin_single_sample_before_LE}) (which hold for \emph{any} symmetric matrix) imply that $\Theta_{i} v(P) = I_{\mathcal{B}(x,x)} v(S)$, where $S \in \mathbb{R}^{n \times n}$, $S = S^{T}$ is given by
\begin{align}
	S 
	=
	A_{i}^{T} P + P A_{i}.
	\label{eq:lem_Thetai_LE}
\end{align}

\noindent (\ref{eq:lem_Thetai_LE}) is an ALE, which since $S = S^{T}$ and since $A_{i} = A - B K_{i}$ is Hurwitz has the unique solution $P = \int_{0}^{\infty} e^{A_{i}^{T} \tau} (-S) e^{A_{i} \tau} \, d\tau$ \cite{AAR_multivariable:book}. Meanwhile, by full column rank of $I_{\mathcal{B}(x,x)}$, that $I_{\mathcal{B}(x,x)} v(S) = 0$ implies $v(S) = 0$, or $S = 0$. Since $S = 0$, we see $P = 0$, whence $v(P) = 0$. Altogether, we have shown that $\Theta_{i}$ has a trivial right null space, so it has full column rank.
$\hfill\blacksquare$

%
%

\begin{theorem}[Equivalence of EIRL Algorithm and Kleinman's Algorithm]\label{thm:IRL_nonlin_Kleinman_equivalence}
Suppose that $l \in \mathbb{N}$ and the sample instants $\{t_{k}\}_{k=0}^{l}$ are chosen such that the matrix $I_{\mathcal{B}(x,x)}$ (\ref{eq:I_Bxy_def}) has full column rank $\underline{n}$. If $K_{0}$ is stabilizing, then the EIRL algorithm and Kleinman's algorithm are equivalent in that the sequences $\{P_{i}\}_{i=0}^{\infty}$ and $\{K_{i}\}_{i=1}^{\infty}$ produced by both are identical. Thus, the stability/convergence conclusions of Theorem \ref{thm:Kleinman_AREs} hold for the EIRL algorithm as well.

\end{theorem}

%
%
\textit{Proof:} 
Follows by induction on $i$, after application of Lemmas \ref{lem:IRL_nonlin_Thetai_full_rank} and \ref{lem:IRL_nonlin_Pi_lsq}.
$\hfill\blacksquare$

%
%

\begin{remark}[EIRL: Reduction in Controller Structural Complexity, Exact Agreement with Classical LQR]\label{rk:controller_structure}
When performing real-world convergence/failure diagnosis, it is paramount to have 1) a concrete structural understanding, and 2) a sound numerical estimate of what controller an algorithm ``should" converge to. Addressing these needs directly: 1) The EIRL algorithm outputs a sequence of linear controllers $\{K_{i}\}_{i=1}^{\infty}$ which are structurally-identical to the optimal LQR controller $K^{*}$ (\ref{eq:Kstar_LQR}), and indeed numerically-identical to the sequence $\{K_{i}\}_{i=1}^{\infty}$ output by Kleinman's algorithm (cf. Theorem \ref{thm:IRL_nonlin_Kleinman_equivalence}). 2) Exact drift dynamics knowledge $f$, $A$ may not be available, but a nominal model $\tilde{f}$, $\tilde{A}$ can be used to instantly estimate the optimal controller $K^{*}$ expected at the output of EIRL. 
Such systematic comparison capabilities are not offered by the existing CT-RL suite \cite{BA_Wallace_J_Si_CT_RL_review:2022}. Indeed, by contrast, for most nonlinear CT-RL algorithms the optimal value and policy are not available in closed-form. Furthermore, the structure of the policy approximations output is entirely dependent on the basis functions selected (generally nonlinear, highly complex, and difficult to troubleshoot). As a result, it is difficult for CT-RL designers to gauge learning performance, which inevitably degrades synthesis to a ``shot-in-the-dark" endeavor.

\end{remark}

%
%

Entirely analogous versions of Lemmas \ref{lem:IRL_nonlin_Pi_lsq} and \ref{lem:IRL_nonlin_Thetai_full_rank} hold for the dEIRL algorithm developed in Section \ref{ssec:dirl_forumlation}, and hence the results of Theorem \ref{thm:IRL_nonlin_Kleinman_equivalence} are preserved by the dEIRL algorithm:

\begin{theorem}[Equivalence of dEIRL Algorithm and Kleinman's Algorithm]\label{thm:DIRL_nonlin_Kleinman_equivalence}
Suppose for $1 \leq j \leq N$ that $l_{j} \in \mathbb{N}$ and the sample instants $\{t_{k,j}\}_{k=0}^{l_{j}}$ are chosen such that the matrix $I_{\mathcal{B}(x_{j},x_{j})}$ (\ref{eq:I_Bxy_def}) has full column rank $\underline{n}_{j}$. If $K_{0,j}$ is such that $A_{jj} - B_{jj} K_{0,j}$ is Hurwitz, then the dEIRL algorithm and Kleinman's algorithm are equivalent in that the sequences $\{P_{i,j}\}_{i=0}^{\infty}$ and $\{K_{i,j}\}_{i=1}^{\infty}$ produced by both are identical. Thus, the stability/convergence conclusions of Theorem \ref{thm:Kleinman_AREs} hold for the dEIRL algorithm as well.

\end{theorem}

%
%

\begin{remark}[dEIRL Algorithm: Decentralized Learning, with or without Dynamic Coupling]\label{rk:DIRL_lin_propts}
The dEIRL algorithm (via Theorem \ref{thm:DIRL_nonlin_Kleinman_equivalence}) guarantees convergence to the solution of the \emph{linear} quadratic regulator problem associated with loop $j$: $(A_{jj}, B_{jj}, Q_{j}, R_{j})$ from state trajectory data generated by the \emph{nonlinear} system $(f, g)$ (\ref{eq:sys_nonlin_2x2}), \emph{regardless} of if $(f, g)$ is dynamically coupled between loops $j = 1, \dots, N$. 
Note that Theorem \ref{thm:DIRL_nonlin_Kleinman_equivalence} involves only a fixed single loop $1 \leq j \leq N$, both in terms of assumptions and results. We in particular call attention to the crux of the hypotheses required in Theorem \ref{thm:DIRL_nonlin_Kleinman_equivalence}: full-column rank of the matrix $I_{\mathcal{B}(x_{j},x_{j})} \in \mathbb{R}^{l_{j} \times \underline{n}_{j}}$ (\ref{eq:I_Bxy_def}). This matrix places requirements on the quality of state trajectory data $x_{j}$ in loop $j$ only. Thus, the dEIRL algorithm is truly decentralized: The loops $j = 1, \dots, N$ may be updated entirely independently, and the designer may focus on data quality in the individual loops rather than for the aggregate system. 
As a result, the dEIRL algorithm offers significant real-world benefits to designers in terms of dimensionality, allowing a single higher-dimensional problem to be partitioned into lower-dimensional subproblems.

\end{remark}



%
%

%
%


%
%

\section{Evaluation Studies}\label{sec:ES}

%
%

The HSV model used in this study was developed in \cite{Marrison_Stengel_HSV_tracking_robust:1998,Wang_Stengel_HSV_tracking_robust:2000} based on NASA Langley's winged-cone tabular aeropropulsive data \cite{Shaughnessy_Pinckney_McMinn_Cruz_Kelley_HSV_modeling_NASA_Langley:1990}. The model in \cite{Marrison_Stengel_HSV_tracking_robust:1998,Wang_Stengel_HSV_tracking_robust:2000} has served as a standard testbed for HSV control development, later being used in seminal works such as \cite{Xu_Mirmirani_Ioannou_HSV_tracking_neural_adaptive:2003,Xu_Mirmirani_Ioannou_HSV_tracking_adaptive_sliding_mode:2004}. The model presented here is identical to that of \cite{Wang_Stengel_HSV_tracking_robust:2000}, with two exceptions: First, we added the elevator-lift increment coefficient $C_{L,\delta_{E}}$ (\ref{eq:C_L_dE}) from data in \cite{Shaughnessy_Pinckney_McMinn_Cruz_Kelley_HSV_modeling_NASA_Langley:1990} to capture nonminimum phase behavior. Second, we have removed angle of attack (AOA) dependence from the thrust coefficient term $k$ (\ref{eq:kx}) which was present in \cite{Marrison_Stengel_HSV_tracking_robust:1998,Wang_Stengel_HSV_tracking_robust:2000}, as thrust coefficient AOA dependencies were considered negligible in the original propulsion model \cite[pp. 12]{Shaughnessy_Pinckney_McMinn_Cruz_Kelley_HSV_modeling_NASA_Langley:1990}, and it was removed in subsequent studies \cite{Xu_Mirmirani_Ioannou_HSV_tracking_neural_adaptive:2003,Xu_Mirmirani_Ioannou_HSV_tracking_adaptive_sliding_mode:2004}.

Instability and nonminimum phase behavior impose respective min/max requirements on closed-loop bandwidth, the combination of which make the HSV a formidable design challenge even for classical methods \cite{Stein_respect_the_unstable:2003}. With the additional obstacles of dimensionality, approximation, and numerics facing CT-RL algorithms, this example is significant and far from a toy problem.

These evaluations were performed in MATLAB R2021a, on an NVIDIA RTX 2060, Intel i7 (9th Gen) processor. All numerical integrations in this work are performed in MATLAB's adaptive \texttt{ode45} solver to ensure solution accuracy.
All code developed for this work can be found at \cite{BA_Wallace_github_TNNLS_2023:webpage}.

%
%

\subsection{Setup, Hyperparameter Selection}\label{ssec:ES_setup}

%
%

%

%
%

\noindent\textbf{HSV Longitudinal Model.}
Consider the following HSV longitudinal model \cite{Marrison_Stengel_HSV_tracking_robust:1998,Wang_Stengel_HSV_tracking_robust:2000, Shaughnessy_Pinckney_McMinn_Cruz_Kelley_HSV_modeling_NASA_Langley:1990}
\begin{align}
	\dot{V} 
	&=
	\frac{T \cos \alpha - D}{m} - \frac{\mu \sin \gamma}{r^{2}},
	\nonumber
	\\
	\dot{\gamma}
	&=
	\frac{L + T \sin \alpha}{m V} - \frac{(\mu - V^{2} r) \cos \gamma}{V r^{2}},
	\nonumber
	\\
	\dot{\theta}
	&=
	q,
	\nonumber
	\\
	\dot{q}
	&=
	\frac{\mathcal{M}}{I_{yy}},
	\nonumber
	\\
	\dot{h}
	&=
	V \sin \gamma,	
	\label{eq:HSV_longitudinal_eqns}
\end{align}

\noindent where $V$ is the vehicle airspeed, $\gamma$ is the flightpath angle (FPA), $\alpha$ is the angle of attack (AOA), $\theta \triangleq \alpha + \gamma$ is the pitch attitude, $q$ is the pitch rate, and $h$ is the vehicle altitude. Here $r(h) = h + R_{E}$ is the total distance from the earth's center to the vehicle, $R_{E} = 20,903,500$ ft is the radius of the earth, and $\mu = G m_{E} = 1.39 \times 10^{16}$ ft$^{3}$/s$^{2}$, where $G$ is Newton's gravitational constant and $m_{E}$ is the mass of the earth. $L, D, T, \mathcal{M}$ are the lift, drag, thrust, and pitching moment, respectively, and are given by
\begin{align}
	L 
	&=
	\frac{1}{2} \rho V^{2} S C_{L},
	&
	D
	&=
	\frac{1}{2} \rho V^{2} S C_{D},		
	\\
	T
	&=
	\frac{1}{2} \rho V^{2} S C_{T},	
	&
	\mathcal{M}
	&=
	\frac{1}{2} \rho V^{2} S \overline{c} C_{M},							
\end{align}

\noindent where $\rho$ is the local air density, $S = 3603$ ft$^{2}$ is the wing planform area, and $\overline{c} = 80$ ft is the mean aerodynamic chord of the wing. Air density $\rho$ and speed of sound $a$ are modeled as functions of altitude $h$ by
\begin{align}
	\rho 
	&=
	0.00238 e^{- \frac{h}{24,000}},
	\\
	a
	&=
	8.99 \times 10^{-9} h^{2} - 9.16 \times 10^{-4} h + 996,
\end{align}
	
\noindent and Mach number $M \triangleq \frac{V}{a}$. The lift, drag, thrust, and pitching moment coefficients are given by
\begin{align}
	C_{L}
	&=
	C_{L,\alpha} + C_{L,\delta_{E}},
	\\
	C_{L,\alpha}
	&=
	\nu \; \alpha \left( 0.493 + \frac{1.91}{M} \right),
	\label{eq:C_L_a}
	\\
	C_{L,\delta_{E}}
	&=
	\left( -0.2356 \alpha^{2} - 0.004518 \alpha - 0.02913 \right) \delta_{E},
	\label{eq:C_L_dE}
	\\
	C_{D}
	&=
	0.0082 \left( 171 \alpha^{2} + 1.15 \alpha + 1 \right) 
	\nonumber
    \\
	&\qquad\qquad
	\times \left( 0.0012 M^{2} - 0.054 M + 1 \right),
%
    \\
	C_{M}
	&=
	C_{M,\alpha} + C_{M,q} + C_{M,\delta_{E}},
	\\
	C_{M,\alpha}
	&=
	10^{-4} \left( 0.06 - e^{-\frac{M}{3}} \right) \left( -6565 \alpha^{2} + 6875 \alpha + 1 \right),
	\\
	C_{M,q}
	&=
	\left( \frac{q \overline{c}}{2 V} \right) \left( - 0.025 M + 1.37 \right)
	\nonumber
	\\
	&\qquad\qquad
	 \times \left( -6.83 \alpha^{2} + 0.303 \alpha - 0.23 \right),
	\\
	C_{M,\delta_{E}}
	&=
	0.0292 (\delta_{E} - \alpha),
\end{align}
\begin{align}
	k 
	&=
	0.0105 \left( 1 + \frac{17}{M} \right),
	\label{eq:kx}
	\\
	C_{T}
	&=
	\begin{cases}
		k (1 + 0.15) \delta_{T}, & \delta_{T} < 1
		\\
		k (1 + 0.15 \delta_{T}), & \delta_{T} \geq 1,
	\end{cases}
\end{align}								

\noindent where $\delta_{E}$ is the elevator deflection, $\delta_{T}$ is the throttle setting, and $\nu \in \mathbb{R}$ (\ref{eq:C_L_a}) is an unknown parameter (nominally 1) representing modeling error in the basic lift increment coefficient $C_{L,\alpha}$. The system (\ref{eq:HSV_longitudinal_eqns}) is fifth-order, with states $x = \left[ V, \, \gamma, \, \theta, \, q, \, h \right]^{T}$. The controls are $u = \left[ \delta_{T}, \, \delta_{E} \right]^{T}$, and we examine the outputs $y = \left[ V, \, \gamma \right]^{T}$. As in \cite{Wang_Stengel_HSV_tracking_robust:2000}, we study a steady level flight cruise condition $q_{e} = 0, \gamma_{e} = 0^{\circ}$, at $M_{e} = 15$, $h_{e} = 110,000$ ft, which corresponds to an equilibrium airspeed $V_{e} = 15,060$ ft/s. At this flight condition, the vehicle is trimmed at $\alpha_{e} = 1.7704^{\circ}$ by the controls $\delta_{T,e} = 0.1756$ ($T_{e} = 4.4966 \times 10^{4}$ lb), $\delta_{E,e} = -0.3947^{\circ}$.

%
%

\begin{remark}[HSV Dynamical Challenges: Instability, Nonminimum Phase, Model Uncertainty]\label{rk:HSV_design_challenges}
The HSV model studied here encompasses a variety of dynamical challenges facing real-world control designers. Firstly, the HSV is open-loop unstable. Linearization of the model about the equilibrium flight condition $(x_{e}, u_{e})$ has open-loop eigenvalues at $s = -0.8291, 0.7165$ (short-period modes), $s = -0.00001 \pm 0.0276 j$ (phugoid modes), and $s = 0.0005$ (altitude mode). The dominant unstable short-period right half plane pole (RHPP) at $s = 0.7165$ is associated with the vehicle pitch-up instability (long vehicle forebody, aftward-set center of mass). As is commonplace with tail-controlled aircraft, the elevator-FPA map is nonminimum phase \cite{Bolender_Doman_HSV_modeling:2005}. The linearized plant has transmission zeros at $s = 8.3938, -8.4620$, the right half plane zero (RHPZ) at $s = 8.3938$ being attributable to the elevator-FPA map (negative lift increment in response to pitch-up elevator deflections). 

Reducing the lift coefficient $\nu < 1$ (\ref{eq:C_L_a}) represents degraded lift efficiency and a more difficult vehicle to control dynamically. The evaluations of Section \ref{ssec:ES_CL_error} study dEIRL learning performance in the presence of a 10\% modeling error $\nu = 0.9$ and a 25\% modeling error $\nu = 0.75$. For perspective, at $\nu = 0.9$, the system has its dominant RHPP at $s = 0.7011$ and RHPZ at $s = 7.9619$, and at $\nu = 0.75$, the system has its dominant RHPP at $s = 0.6681$ and RHPZ at $s = 7.2664$. Thus, the pole/zero ratio drops from 11.72 nominally ($\nu = 1$), to 11.36 ($\nu = 0.9$), to 10.88 ($\nu = 0.75$). Aerodynamic modeling errors are quite common in aerospace applications, especially in the uniquely-challenging HSV context \cite{McRuer_HSV_modeling:1991,Schmidt_HSV_modeling:1992,Bolender_Doman_HSV_modeling:2005,Williams_Bolender_Doman_Morataya_HSV_modeling:2006,AA_Rodriguez_Dickeson_etal_HSV:2008,Dickeson_AA_Rodriguez_Sridharan_etal_HSV:2009}.
Between aeropropulsive modeling errors in the tabular data and curve fitting errors, a 10\% error in lift coefficient is to be expected. A 25\% error is severe, chosen deliberately so to push the learning limits of dEIRL.

\end{remark}

%
%

\noindent\textbf{Decentralized Design Framework.} 
This work implements a decentralized design methodology largely inspired by the framework developed in \cite{Dickeson_AA_Rodriguez_Sridharan_etal_HSV_decentralized_control:2009}, wherein controllers are designed separately for the weakly-coupled velocity subsystem (associated with the airspeed $V$ and throttle control $\delta_{T}$) and rotational subsystem (associated with the FPA $\gamma$, attitude $\theta, q$, and elevator control $\delta_{E}$). As in \cite{Dickeson_AA_Rodriguez_Sridharan_etal_HSV_decentralized_control:2009}, for controllability reasons we do not feed back altitude $h$ in the control design, though altitude is still included in the nonlinear simulation. In order to achieve zero steady-state error to step reference commands, we augment the plant at the output with the integrator bank $z = \int y \, d\tau = \left[ z_{V}, \, z_{\gamma}  \right]^{T} = \left[ \int V \, d\tau, \, \int \gamma \, d\tau \right]^{T}$. For dEIRL, the state/control vectors are thus partitioned as $x_{1} = \left[ z_{V}, \, V \right]^{T}$, $u_{1} = \delta_{T}$ ($n_{1} = 2$, $m_{1} = 1$) and $x_{2} = \left[ z_{\gamma}, \, \gamma, \, \theta, \, q \right]^{T}$, $u_{2} = \delta_{E}$ ($n_{2} = 4$, $m_{2} = 1$). Applying the LQ servo design framework \cite{AAR_multivariable:book} to each of the loops yields a proportional-integral (PI) velocity controller and a proportional-derivative (PD)/PI inner/outer FPA controller structurally identical to those presented in \cite{Dickeson_AA_Rodriguez_Sridharan_etal_HSV_decentralized_control:2009}. It is these optimal LQ controller parameters which the proposed methods will learn online.

%
%

\noindent\textbf{Hyperparameter Selection.} 
We next discuss hyperparameter selections for the learning algorithms. For sake of comparison, we hold all hyperparameter selections constant across Evaluations 1 and 2. First, we discuss cost structure.
In the velocity loop $j = 1$, we choose the state penalty $Q_{1} = I_{2}$ and control penalty $R_{1} = 15$, while in the FPA loop $j = 2$ we choose the state penalty $Q_{2} = \texttt{diag}(1, 1, 0, 0)$ and control penalty $R_{2} = 0.01$. These cost structure parameters were selected to yield optimal LQ designs $K_{1}^{*}$ (\ref{eq:ES_nom_K1star}), $K_{2}^{*}$ (\ref{eq:ES_nom_K2star}), which achieve closed-loop step response specifications comparable to previous works \cite{Dickeson_AA_Rodriguez_Sridharan_etal_HSV_decentralized_control:2009,Dickeson_AA_Rodriguez_Sridharan_etal_HSV:2009,Dickeson_PhD_thesis_ASU:2012}: A 90\% rise time in velocity $t_{r, V, 90 \%} = 31.99$ s and FPA $t_{r, \gamma, 90 \%} = 4.56$ s, a 1\% settling time in velocity $t_{s, V, 1 \%} = 78.18$ s and FPA $t_{s, \gamma, 1 \%} = 8.643$ s, percent overshoot in velocity $M_{p, V} =  4.24$\% and FPA $M_{p, \gamma} = 3.988$\%. Next, using the decentralized control method described in \cite{Dickeson_AA_Rodriguez_Sridharan_etal_HSV_decentralized_control:2009,Dickeson_PhD_thesis_ASU:2012} (which performs decentralized LQR designs on simplified versions of the diagonal plant terms $P_{jj}$), 
we arrive at the following initial stabilizing controllers
\begin{align}	
	K_{0,1}
	&=
	\left[
	\begin{array}{cc}
		0.2582 &   4.3570
	\end{array}
	\right],
	\label{eq:ES_K01}
	\\
	K_{0,2}
	&=
	\left[
	\begin{array}{cccc}
		10.0000 &  26.3299 &   1.6501  &  1.0124
	\end{array}
	\right],
	\label{eq:ES_K02}
	\\
	K_{0}
	&=
	\left[
	\begin{array}{cc}
		K_{0,1} & 0
		\\
		0 & K_{0,2}
	\end{array}
	\right].
	\label{eq:ES_K0}
\end{align}

Next, we discuss the injected signals. For the exploration noise $d$ (used by all methods except the original IRL formulation \cite{Vrabie_Lewis_IRL:2009}) we choose $d_{1}(t) = 0.1 \sin \left( \frac{2\pi}{25} t\right) + 0.1 \sin \left( \frac{2\pi}{250} t\right) + 0.2$ and $d_{2}(t) = 10 \sin \left( \frac{2\pi}{6} t\right) + 5 \cos \left( \frac{2\pi}{50} t\right) + 2.5 \sin \left( \frac{2\pi}{25} t \right)$. As a brief aside, In light of the discussion in Section \ref{ssec:multi_injection}, it is no coincidence for the noise $d_{2}$ that we have placed the dominant term at the frequency $\omega = \frac{2\pi}{6} \approx 1$ rad/s -- the peak $P$-sensitivity $T_{d_{i} y}$ frequency (cf. Figure \ref{fig:PSu}). This choice maximizes the excitation efficiency. For the reference command $r$ (used in the MI mode only), we choose $r_{1}(t) = 10 \cos \left( \frac{2\pi}{10} t\right) + 10 \sin \left( \frac{2\pi}{25} t\right) + 50 \sin \left( \frac{2\pi}{200} t\right)$ and $r_{2}(t) = 0.02 \cos \left( \frac{2\pi}{3} t\right) + 0.1 \sin \left( \frac{2\pi}{6} t\right) + 0.25 \sin \left( \frac{2\pi}{15} t\right)$. 

We now discuss learning hyperparameters (i.e., sample period $T_{s} = t_{k} - t_{k-1}$, number of samples collected $l$, number of iterations $i^{*}$). Selections for all methods, summarized in Table \ref{tb:learning_hyperparams}, reflect two objectives: The first to keep hyperparameter selections constant across methods wherever possible, and the second to vary parameter selections when they result in significant numerical conditioning improvements in order to present a ``best-case" illustration. 

%
%

\begin{table}[h]
	\caption{Learning Hyperparameters}
	\centering
	\begin{tabular}{|c|c||c|c|c|}
	        \hline
	        Algorithm & Loop $j$ & $T_{s,j}$ (s) & $l_{j}$ & $i^{*}_{j}$  
	        \\
	        \hline
	        \hline
	        IRL (old) \cite{Vrabie_Lewis_IRL:2009} & 1 &  0.15  & 25  & 5   
	        \\
	        \hline	      
	        \hline
	        EIRL & 1 &  5  & 25  & 5   
	        \\
	        \hline		          
	        \hline
	        \multirow{2}{*}{dEIRL} & 1 & 6 & 15 & 5
	        \\
	        \hhline{|~|-||-|-|-|}
			& 2 & 2 & 25 & 5   
	        \\
	        \hline	  
	\end{tabular}
	\label{tb:learning_hyperparams}
\end{table}

The latter objective surfaces particularly in selection of the sample period $T_{s}$. As has been systematically illustrated in the CT-RL comparative study \cite{BA_Wallace_J_Si_CT_RL_review:2022}, a relatively short sample period $T_{s} = 0.15$ s is required for the original IRL algorithm \cite{Vrabie_Lewis_IRL:2009} due to the lack of ability to insert probing noise $d$ (see Section \ref{ssec:ES_nom} for further discussion). By contrast, for dEIRL the designer can select the sample period $T_{s,j}$ in accordance with the natural bandwidth of the dynamics in the respective loop $j$ -- a crucial numerical advantage of decentralization. In this example, the complementary sensitivity response $T_{r y}$ in Figure \ref{fig:Te} shows that the velocity $j = 1$ and FPA $j = 2$ loops are separated by a decade in bandwidth. As such, using a single sample period to capture the dynamical features of both of these loops is naturally troublesome. Reaffirming these intuitions experimentally, a longer sample period $T_{s,1} = 6$ s is observed numerically-favorable in the lower-bandwidth velocity loop, while a shorter sample period $T_{s,2} = 2$ s is favorable in the higher-bandwidth FPA loop. In the case of EIRL, the designer is afforded the luxury of probing noise injection (and/or reference command injection in the MI case), but is still required to choose a single sample period $T_{s}$. It is then intuitive that a sample period $T_{s} = 5$ s (i.e., between the favorable dEIRL selections $T_{s,2} = 2$ s and $T_{s,1} = 6$ s) is observed to yield the best conditioning for EIRL.

For selection of the number of data points $l$, we note that this system has regression dimensions $\underline{n} = 21$, $\underline{n}_{1} = 3$, $\underline{n}_{2} = 10$, which serve as lower bounds for the number of samples collected in the respective regression. The lower dimensionality of the velocity loop $j = 1$ allows for fewer data points $l_{1} = 15$ to be collected for dEIRL than for the other methods, which indeed proved numerically favorable. 

Next, we discuss system initial conditions (ICs). Since the original IRL algorithm \cite{Vrabie_Lewis_IRL:2009} does not allow probing noise injection, the only means of excitation available to the designer is initialization of the system away from trim $x_{e}$. To this end, for the old IRL algorithm \cite{Vrabie_Lewis_IRL:2009}, we initialize the HSV at the airspeed $V_{0} = V_{e} + 1000$ ft/s and FPA $\gamma_{0} = \gamma_{e} + 2^{\circ} = 2^{\circ}$, and otherwise all remaining ICs are set to trim. For the algorithms developed in this work, we need not bother with this practice and hence initialize the system at trim $x_{0} = x_{e}$.

%
%
 
\begin{remark}[Original IRL Algorithm \cite{Vrabie_Lewis_IRL:2009} Implementation Notes]\label{rk:IRL_old_implementation}
For the original IRL algorithm \cite{Vrabie_Lewis_IRL:2009}, we use critic basis activation functions $\mathcal{B}(x,x)$ to make its activation functions identical to EIRL. Due to the severe numerical conditioning issues associated with the original IRL algorithm \cite{Vrabie_Lewis_IRL:2009}, the closed-loop system goes unstable upon the first controller update regardless of hyperparameter selections (cf. Section \ref{ssec:ES_nom} for further discussion). In order to address this issue and offer a conditioning analysis of the original IRL algorithm similar to what would be observed had its controller updates been stabilizing, we run it simulating under the initial stabilizing controller $K_{0}$ (\ref{eq:ES_K0}) without updating at each iteration $i$.
\end{remark}

%
%
 
\begin{remark}[Hyperparameter Selection: Now an Intuitive Synthesis Tool]\label{rk:hyperparam_selection_intuitive}
Algorithm design efforts in this work deliberately reduce the hyperparameter selection space in order to simplify practical design. 
In particular, the designer is not confronted by a vast array of obscure tuning gains/matrices, nor are they burdened with the haphazard and numerically-troublesome task of basis function selection -- both practical issues of current CT-RL formulations \cite{BA_Wallace_J_Si_CT_RL_review:2022}. 

\end{remark}



%
%


%
%

%
%
\renewcommand{\relpath}{figures/ES_nom/}

\subsection{Evaluation 1: Conditioning \& Convergence Study on Nominal Model}\label{ssec:ES_nom}

%
%
%
%

\renewcommand{\relpathone}{cond/}

\begin{figure*}[ht]
\begin{minipage}{0.4\textwidth}
    \centering
    \includegraphics[height=1.2\figsize]{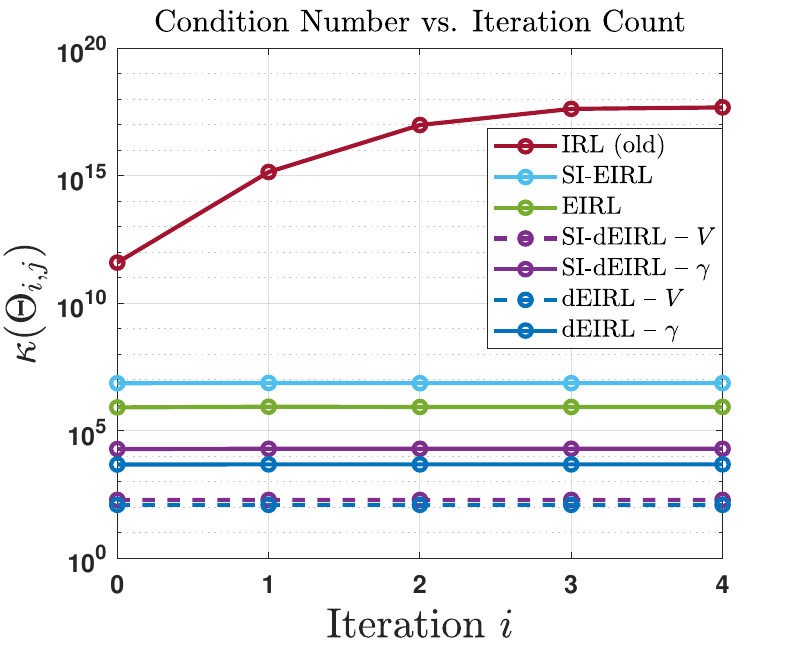}	
    \vspace{-0.75cm}	
    \captionof{figure}{Eval. 1: Condition number versus iteration count $i$.}
	\label{fig:ES_nom_cond_A_vs_i}	
\end{minipage}%
\hfill
\begin{minipage}{0.6\textwidth}
      \captionof{table}{Eval. 1: Max, Min Conditioning}
	\centering
    \small
	\begin{tabular}{|c|c||c|c|c|c|}
	        \hline
	        Algorithm & Loop $j$ & $\max\limits_{i}(\kappa(\Theta_{i,j}))$ & $\min\limits_{i}(\kappa(\Theta_{i,j}))$ & $i_{\max\kappa}$  & $i_{\min\kappa}$
	        \\
	        \hline
	        \hline
	        IRL (old) & 1 & \cellcolor{black!\darkshade} 5.00e+17 & \cellcolor{black!\darkshade} 3.97e+11 & 4 &  0
	        \\
	        \hline	      
	        \hline
	        SI-EIRL & 1 & \cellcolor{black!\lightshade} 7.52e+06 & \cellcolor{black!\lightshade} 7.34e+06 & 1 & 0
	        \\
	        \hline	
	        \hline
	        EIRL & 1 & \cellcolor{black!\lightshade} 8.79e+05 & \cellcolor{black!\lightshade} 8.27e+05 & 1 & 0
	        \\
	        \hline		 	        	          
	        \hline
	        \multirow{2}{*}{SI-dEIRL} & 1 & 193.19 & 193.16 & 0 & 1
	        \\
	        \hhline{|~|-||-|-|-|-|}
			& 2 & 1.97e+04 & 1.93e+04 & 1 & 0
	        \\
	        \hline	  
	        \hline
	        \multirow{2}{*}{dEIRL} & 1 & 123.26 & 123.25 & 0 & 1
	        \\
	        \hhline{|~|-||-|-|-|-|}
			& 2 & 4.81e+03 & 4.74e+03 & 1 & 0
	        \\
	        \hline	  	        
	\end{tabular}
	\label{tb:ES_nom_min_max_cond}
\end{minipage}
\end{figure*}

In this section, we demonstrate conditioning and convergence effectiveness \cite{BA_Wallace_J_Si_CT_RL_review:2022} of EIRL and dEIRL.
%
%
Specifically, we first show the significant conditioning reduction moving from the original IRL formulation \cite{Vrabie_Lewis_IRL:2009} to any of the methods developed here. Second, we illustrate the successive conditioning improvements associated with multi-injection (SI $\rightarrow$ MI) and decentralization (EIRL $\rightarrow$ dEIRL). 
As a convention, we include the prefix SI- (e.g., SI-EIRL) to make explicit when an algorithm is in SI mode; otherwise, without a prefix the algorithm is assumed to be in MI mode (e.g., EIRL).
For this comparative study, we consider the nominal HSV model described in Section \ref{ssec:ES_setup} (i.e., zero lift-coefficient modeling error $\nu = 1$ (\ref{eq:C_L_a})). 

\begin{figure*}[ht]
    \centering
    \subfloat[]{\includegraphics[height=\figsize]{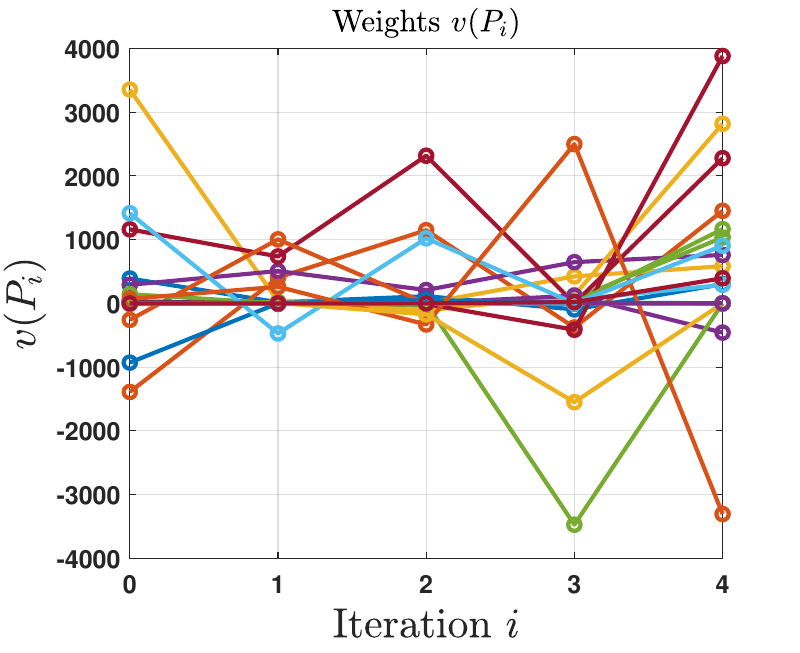}%
    \label{fig:ES_nom_vPi_irl_old}}
    \hfil
    \subfloat[]{\includegraphics[height=\figsize]{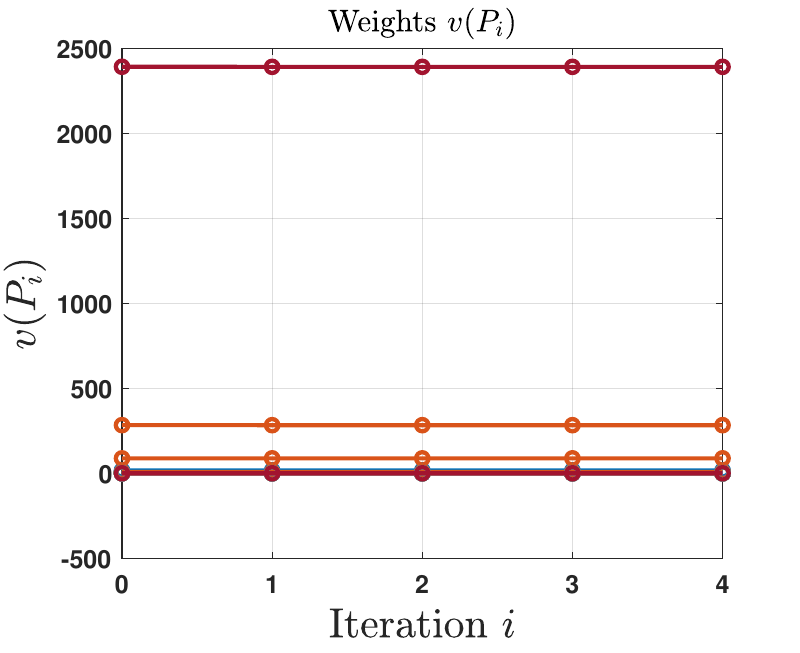}%
    \label{fig:ES_nom_vPi_si_irl}}
    \caption{Eval. 1: Weight responses $v(P_{i})$ (\ref{eq:IRL_nonlin_lsq}). 
    (a): Original IRL algorithm \cite{Vrabie_Lewis_IRL:2009}. 
    (b): SI-EIRL (Section \ref{ssec:irl_forumlation}).}
    \label{fig:ES_nom_vPi}
\end{figure*}

Before we begin, we motivate why conditioning is integral to the study of CT-RL algorithm learning performance \cite{BA_Wallace_J_Si_CT_RL_review:2022}.
Existing ADP-based CT-RL algorithms which use linear regressions in the form of Equations (\ref{eq:IRL_nonlin_lsq}) and (\ref{eq:DIRL_nonlin_lsq}) to yield their NN weights exhibit condition numbers on the order of $\kappa(\Theta) \approx 10^{5} - 10^{11}$ for simple second-order academic examples \cite[Table VII]{BA_Wallace_J_Si_CT_RL_review:2022}. The original IRL algorithm \cite{Vrabie_Lewis_IRL:2009} has conditioning of $\kappa(\Theta_{i}) = 5 \times 10^{17}$ in this HSV study (cf. Table \ref{tb:ES_nom_min_max_cond}). It is then to be expected that IRL's NN weights exhibit large oscillations on the order of 8,000 and fail to converge (cf. Figure \ref{fig:ES_nom_vPi_irl_old}). The culprit stifling the learning performance of IRL (and of the existing ADP-based CT-RL suite, more broadly \cite{BA_Wallace_J_Si_CT_RL_review:2022}) is poor conditioning. It is thus vital that future works show rigorous numerical conditioning studies when proving new CT-RL algorithms.

%
%

\noindent\textbf{\ul{Conditioning Analysis.}}
The entirety of this analysis is summarized in Figure \ref{fig:ES_nom_cond_A_vs_i} and Table \ref{tb:ES_nom_min_max_cond}. Figure \ref{fig:ES_nom_cond_A_vs_i} displays the iteration-wise condition number of each of the methods studied (i.e., of the matrix $\Theta_{i}$ (\ref{eq:IRL_nonlin_lsq_A}) for IRL \cite{Vrabie_Lewis_IRL:2009} and EIRL, and $\Theta_{i,j}, \, j = 1, 2$ (\ref{eq:DIRL_nonlin_lsq_A}) for dEIRL). Table \ref{tb:ES_nom_min_max_cond} displays the maximum and minimum of the conditioning data presented in Figure \ref{fig:ES_nom_cond_A_vs_i} (i.e., taken over iteration $0 \leq i \leq i^{*}-1$), alongside the iteration of max conditioning $i_{\max\kappa}$ and min conditioning $i_{\min\kappa}$. 
%
%
Figure \ref{fig:ES_nom_cond_A_vs_i} reveals that the original IRL algorithm \cite{Vrabie_Lewis_IRL:2009} suffers from the worst conditioning of the methods presented, featuring a monotonic increase from $4 \times 10^{11}$ at $i = 0$ to $5 \times 10^{17}$ at $i = 4$ (Table \ref{tb:ES_nom_min_max_cond}). Conditioning on this order in a monotonic degradation pattern has been demonstrated by IRL previously and is associated with PE issues as the system approaches the origin under the stabilizing policy $K_{0}$ without probing noise excitation (cf. \cite[Section \CTRLSecCSI]{BA_Wallace_J_Si_CT_RL_review:2022} for in-depth analysis). By contrast, SI-EIRL (which naturally possesses the worst conditioning properties of the methods developed here, by virtue of not taking advantage of MI or decentralization) features steady conditioning on the order of $7.5 \times 10^{6}$ (Table \ref{tb:ES_nom_min_max_cond}). This represents an eleven order of magnitude improvement in worst-case conditioning over prior ADP methods.

%
%
%
\begin{remark}[Multi-Injection: An Order of Magnitude Reduction in Conditioning]
Moving down Table \ref{tb:ES_nom_min_max_cond}, the effect of MI is generally an order of magnitude conditioning reduction. Conditioning of SI-EIRL is on the order of $7.5 \times 10^{6}$, and EIRL $8.5 \times 10^{5}$. SI-dEIRL in loop $j = 2$ has conditioning on the order of $2 \times 10^{4}$, and dEIRL $4.75 \times 10^{3}$ in this loop. The one case where this trend does not hold is in dEIRL loop $j = 1$, wherein SI-dEIRL conditioning is 193 and dEIRL 123. We attribute the less dramatic reduction to diminishing returns associated with the already-favorable conditioning in this loop, although a reduction of 36\% is still significant.

\end{remark}

%
%
%
\begin{remark}[Decentralization: A Two-to-Four Order of Magnitude Reduction in Conditioning]
Examining Table \ref{tb:ES_nom_min_max_cond} again, we see that decentralization achieves numerical improvements even more impressive than those of MI. Moving from SI-EIRL to its decentralized SI-dEIRL counterpart, conditioning is reduced from $7.5 \times 10^{6}$ to 193 in loop $j = 1$ and $2 \times 10^{4}$ in loop $j = 2$ -- a reduction of four and two orders of magnitude, respectively. The trend is similar from EIRL ($8.5 \times 10^{5}$) to dEIRL ($123$ in loop $j = 1$ and $4.75 \times 10^{3}$ in loop $j = 2$), a respective reduction of three and two orders of magnitude.

\end{remark}

%
%
%
\begin{remark}[Multi-Injection and Decentralization: The Curses of Conditioning and Dimensionality Mollified]
In all, the cumulative numerical improvements from the original IRL formulation \cite{Vrabie_Lewis_IRL:2009} to dEIRL represent momentous reductions in worst-case conditioning (fifteen orders of magnitude in the velocity loop $j = 1$, fourteen in the FPA loop $j = 2$). Moving from SI-EIRL to successive application of MI and decentralization reduces conditioning by four orders of magnitude in loop $j = 1$ and three orders of magnitude in loop $j = 2$. 




\end{remark}

%
%

\renewcommand{\relpathone}{ES_nom_irl_old_training/DIRL/}	
\renewcommand{\relpathtwo}{ES_nom_irl_training/DIRL/}

%
%

\noindent\textbf{\ul{Convergence Analysis.}}
This model has the following optimal LQ controllers
\begin{align}	
	&
	K_{1}^{*}
	=
	\left[
	\begin{array}{cc}
		0.2582  &  4.3577
	\end{array}
	\right],
	\label{eq:ES_nom_K1star}
	\\
	&
	K_{2}^{*}
	=
	\left[
	\begin{array}{cccc}
		10.0000  & 26.3393  &  1.6514  &  0.9921
	\end{array}
	\right],
	\label{eq:ES_nom_K2star}
	\\
	&
	K^{*} =
	\nonumber
	\\
	&
	\hspace{-0.1cm}
	\text{
	{\scriptsize
	$\left[
	\begin{tabular}{cc|cccc}
	    0.2581  &  4.3622  &  0.0074  &  0.0814 &   0.0000  &  0.0001
	    \\
	    \hline
	   -0.2865  & -1.1120  &  9.9959  & 26.3120 &   1.6512  &  0.9921
	\end{tabular}
	\right]$
	}%
	}.
	\label{eq:ES_nom_Kstar}
\end{align}

We have plotted the weight responses $v(P_{i})$ for the original IRL algorithm \cite{Vrabie_Lewis_IRL:2009} in Figure \ref{fig:ES_nom_vPi_irl_old}. Due to poor conditioning (Figure \ref{fig:ES_nom_cond_A_vs_i}, Table \ref{tb:ES_nom_min_max_cond}), the weights update erratically and fail to converge. This same qualitative weight behavior was illustrated systematically in \cite[Figure \CTRLFigIRLWeights]{BA_Wallace_J_Si_CT_RL_review:2022} on a simple second-order academic example (for which conditioning was on the order of $10^{11}$), so it is not surprising that we encounter it for the HSV here. For comparison, we have plotted the corresponding weight responses for SI-EIRL in Figure \ref{fig:ES_nom_vPi_si_irl}. These two methods use the same basis functions, yet the SI-EIRL weights converge nicely in the $i^{*} = 5$ iterations. Since SI-EIRL naturally has the worst conditioning properties of the methods developed in this work, Figure \ref{fig:ES_nom_vPi_si_irl} represents the ``worst-case" weight response of the proposed methods. 

%
%
%
\begin{remark}[All Proposed Methods Deliver Real-World Synthesis Guarantees]
Each of the proposed methods successfully converges to their respective optimal controller $K^{*}$. Indeed, the largest final controller error $\norm{K_{i^{*}} - K^{*}}$ exhibited by any of the methods is only $4.63 \times 10^{-3}$ (by SI-EIRL). For comparison, dEIRL exhibits final controller errors $\norm{K_{i^{*},1} - K_{1}^{*}} = 1.11 \times 10^{-6}$ and $\norm{K_{i^{*},2} - K_{2}^{*}} = 2.70 \times 10^{-5}$. Thus, this evaluation study affirms that the proposed methods achieve real-world convergence performance in exact accordance with their theoretical guarantees (Section \ref{sec:theoretical_results}) -- perhaps a first in ADP-based CT-RL \cite{BA_Wallace_J_Si_CT_RL_review:2022}.

\end{remark}

\renewcommand{\relpath}{figures/ES_CL_error/}

\subsection{Evaluation 2: dEIRL Optimality Recovery Study on Perturbed Models}\label{ssec:ES_CL_error}

Having now demonstrated a systematic framework for improving learning numerics through an in-depth conditioning analysis of the developed algorithms, for this study we hone our focus to the flagship method: dEIRL. Having demonstrated dEIRL's convergence properties on the nominal HSV model $\nu = 1$ (\ref{eq:C_L_a}), we now analyze its convergence when the model is perturbed from the nominal model to $\nu = 0.9$ (a 10\% modeling error) and $\nu = 0.75$ (a 25\% modeling error), resulting in a more challenging control problem (cf. Remark \ref{rk:HSV_design_challenges} for discussion). We ask: How well does dEIRL recover the optimal controller $K_{j}^{*}$ in loop $j$ associated with the true perturbed model $w_{j} = f_{j}(x) - A_{jj} x_{j}$ (\ref{eq:nonlin_sys_dirl_rewritten}) ($\nu \neq 1$), using in its learning regression (\ref{eq:DIRL_nonlin_lsq}) only the \emph{estimated} dynamics $\tilde{w}_{j} = \tilde{f}_{j}(x) - \tilde{A}_{jj} x_{j}$ ($\nu = 1$)?

%
%

\noindent\textbf{\ul{Conditioning Analysis.}}
Before delving into the main convergence results, we first provide a brief conditioning analysis. For $\nu = 0.9$, dEIRL has max conditioning $\max\limits_{i}(\kappa(\Theta_{i,1})) = 111.13$ and $\max\limits_{i}(\kappa(\Theta_{i,2})) = 6.20 \times 10^{3}$, while for $\nu = 0.75$, dEIRL has max conditioning $\max\limits_{i}(\kappa(\Theta_{i,1})) = 90.89$ and $\max\limits_{i}(\kappa(\Theta_{i,2})) = 8.93 \times 10^{3}$. Overall, conditioning has remained largely unchanged in the velocity loop $j = 1$ (cf. Table \ref{tb:ES_nom_min_max_cond}). Even in the higher-dimensional, unstable, nonminimum-phase FPA loop $j = 2$ directly affected by the lift-coefficient modeling error $\nu$ (\ref{eq:C_L_a}), conditioning performance has only degraded slightly.
In sum, dEIRL possesses inherently-favorable conditioning properties which are robust with respect to (even severe) modeling errors -- a vital real-world performance validation.

%
%

\noindent\textbf{\ul{Convergence Analysis.}}
Running dEIRL for $i^{*} = 5$ iterations and $\nu = 0.9$, we have
\begin{align}	
	K_{i^{*},1}
	&=
	\left[
	\begin{array}{cc}
		0.2582  &  4.3579
	\end{array}
	\right],
	\label{eq:ES_CL_error_nu_0p9_Kistar1}
	\\
	K_{1}^{*}
	&=
	\left[
	\begin{array}{cc}
		0.2582  &  4.3580
	\end{array}
	\right],
	\label{eq:ES_CL_error_nu_0p9_K1star}	
	\\	
	K_{i^{*},2}
	&=
	\left[
	\begin{array}{cccc}
		10.0171 &  27.1052  &  1.5828  &  0.9741
	\end{array}
	\right],
	\label{eq:ES_CL_error_nu_0p9_Kistar2}	
	\\
	K_{2}^{*}
	&=
	\left[
	\begin{array}{cccc}
		10.0000  & 27.0327  &  1.5685  &  0.9671
	\end{array}
	\right].
	\label{eq:ES_CL_error_nu_0p9_K2star}	
\end{align}

\noindent For $\nu = 0.75$, we have
\begin{align}	
	K_{i^{*},1}
	&=
	\left[
	\begin{array}{cc}
		0.2582  &  4.3585
	\end{array}
	\right],
	\label{eq:ES_CL_error_nu_0p75_Kistar1}
	\\
	K_{1}^{*}
	&=
	\left[
	\begin{array}{cc}
		0.2582  &  4.3586
	\end{array}
	\right],
	\label{eq:ES_CL_error_nu_0p75_K1star}	
	\\	
	K_{i^{*},2}
	&=
	\left[
	\begin{array}{cccc}
		10.0397  & 28.5706   & 1.4712  &  0.9539
	\end{array}
	\right],
	\label{eq:ES_CL_error_nu_0p75_Kistar2}	
	\\
	K_{2}^{*}
	&=
	\left[
	\begin{array}{cccc}
		10.0000 &  28.2496  &  1.4303   & 0.9238
	\end{array}
	\right].
	\label{eq:ES_CL_error_nu_0p75_K2star}	
\end{align}

\noindent In Table \ref{tb:ES_CL_error_final_K_error}, we show the controller error reduction $\norm{K_{i,j} - K_{j}^{*}}$ from the initial controllers $K_{0,1}$ (\ref{eq:ES_K01}), $K_{0,2}$ (\ref{eq:ES_K02}) (i.e., the nominal decentralized LQR controllers \cite{Dickeson_AA_Rodriguez_Sridharan_etal_HSV_decentralized_control:2009,Dickeson_PhD_thesis_ASU:2012}) to the final controllers $K_{i^{*},1}$ (\ref{eq:ES_CL_error_nu_0p9_Kistar1}), (\ref{eq:ES_CL_error_nu_0p75_Kistar1}) and $K_{i^{*},2}$ (\ref{eq:ES_CL_error_nu_0p9_Kistar2}), (\ref{eq:ES_CL_error_nu_0p75_Kistar2}), respectively.

%
%

\begin{table}[h]
	\caption{Eval. 2: dEIRL Optimality Recovery}
	\begin{minipage}{0.5\textwidth}	
	\centering
	\begin{tabular}{|c|c||c|c|c|}
	        \hline
	        \multirow{2}{*}{$\nu$ (\ref{eq:C_L_a})} & \multirow{2}{*}{Loop $j$} & \multicolumn{2}{c|}{$\norm{K_{i,j} - K_{j}^{*}}$} & \% Reduction
	        \\
	        \hhline{|~|~||-|-|~|}
	        & & $i = 0$ (Nom. LQ) & $i = i^{*}$ & $i = 0 \rightarrow i^{*}$
	        \\
	        \hline 
	        \hline
	        \multirow{2}{*}{0.9} & 1 & 9.10e-04 & 2.30e-05 & 97.5
	        \\
	        \hhline{|~|-||-|-|-|}
			& 2 & 0.709 & 0.0761 & 89.3
	        \\
	        \hline	  	
	        \hline
	        \multirow{2}{*}{0.75} & 1 & 0.00151 & 7.30e-05 & 95.2
	        \\
	        \hhline{|~|-||-|-|-|}
			& 2 & 1.934 & 0.327 & 83.1
	        \\
	        \hline	 	                
	\end{tabular}
	\vspace{-0.3cm}	
	\end{minipage}	
	\label{tb:ES_CL_error_final_K_error}
\end{table}

\begin{remark}[dEIRL Optimality Recovery Capability]
Examining Table \ref{tb:ES_CL_error_final_K_error}, for the 10\% modeling error $\nu = 0.9$, dEIRL reduces controller optimality error by at least an order of magnitude in each loop. For the 25\% modeling error $\nu = 0.75$, dEIRL reduces controller optimality error by at least 80\% in each loop. It is at this point intuitive that dEIRL converges closer to the optimal controllers for the smaller 10\% modeling error than for the severe 25\% modeling error, and that reductions are more significant in the velocity loop $j = 1$ than in the FPA loop $j = 2$.

This capability is one of great practical utility. Previously, if a designer synthesized an initial LQR controller $K_{0,j}$ (i.e., optimal with respect to the nominal linear drift dynamics $\tilde{A}_{jj}$), they had to content themselves with this design and had no real-world means of improvement. Here we demonstrate that, via the estimate $\tilde{w}_{j}$ ($\nu = 1$), dEIRL outputs a controller $K_{i^{*},j}$ significantly closer to the optimal $K_{j}^{*}$ than the original estimate $K_{0,j}$. It is in this sense that dEIRL partially \emph{recovers} optimality of the controller without explicit knowledge of the system dynamics $w_{j}$ ($\nu \neq 1$) (\ref{eq:nonlin_sys_dirl_rewritten}). 

\end{remark}

%
%

\noindent\textbf{\ul{Closed-Loop Performance Analysis.}}
Having illustrated that dEIRL numerically recovers the optimal controllers $K_{j}^{*}$ in each loop, we conclude this section with an analysis of how it recovers optimal closed-loop performance. To this end, we issue a 100 ft/s step-velocity command and a 1$^{\circ}$ step-FPA command to the (nonlinear, coupled) perturbed HSV models, simulating under the nominal LQ controllers $K_{0,j}$ $(\nu = 1)$, dEIRL controllers $K_{i^{*},j}$, and optimal LQ controllers $K_{j}^{*}$ $(\nu \neq 1)$. The resulting closed-loop step response characteristics in each loop $j$ are listed in Table \ref{tb:ES_CL_error_step_resp_characteristics}, including the 90\% rise time $t_{r, y_{j}, 90 \%}$, 1\% settling time $t_{s, y_{j}, 1 \%}$, and percent overshoot $M_{p,y_{j}}$ $(j = 1, 2)$.

%
%

\begin{table}[h]
	\caption{Eval. 2: Closed-Loop Step Response Characteristics}
	\begin{minipage}{0.5\textwidth}	
	\centering
	\begin{tabular}{|c|c|c||c|c|c|}
	        \hline
	        \multirow{2}{*}{$\nu$ (\ref{eq:C_L_a})} & \multirow{2}{*}{Loop $j$} & \multirow{2}{*}{Algorithm} &  $t_{r, y_{j}, 90 \%}$ & $t_{s, y_{j}, 1 \%}$ & $M_{p,y_{j}}$
	        \\
	        \hhline{|~|~|~||~|~|~|}
	        & & & (s) & (s) & (\%)
	        \\
	        \hline   	
	        \hline
	        \multirow{6}{*}{0.9} & \multirow{3}{*}{1} & Nom. LQ & 32.33 & 78.34 & 4.25
	        \\
	        \hhline{|~|~|-||-|-|-|}
			&  & dEIRL & 31.99 & 78.26 & 4.24
	        \\
	        \hhline{|~|~|-||-|-|-|}
			&  & Opt. LQ & 31.94 & 78.27 & 4.24
	        \\	       
	        \hhline{|~|=|=||=|=|=|}
	         & \multirow{3}{*}{2} & Nom. LQ & 4.75 & 9.97 & 7.11
	        \\
	        \hhline{|~|~|-||-|-|-|}
			&  & dEIRL & 4.80 & 9.33 & 5.26
	        \\
	        \hhline{|~|~|-||-|-|-|}
			&  & Opt. LQ & 4.52 & 9.16 & 5.12
	        \\	       	        
	        \hline	  	
	        \hline
	        \multirow{6}{*}{0.75} & \multirow{3}{*}{1} & Nom. LQ & 32.33 & 78.76 & 4.27
	        \\
	        \hhline{|~|~|-||-|-|-|}
			&  & dEIRL & 31.62 & 78.11 & 4.25
	        \\
	        \hhline{|~|~|-||-|-|-|}
			&  & Opt. LQ & 32.17 & 78.41 & 4.25
	        \\	       
	        \hhline{|~|=|=||=|=|=|}
	         & \multirow{3}{*}{2} & Nom. LQ & 5.13 & 16.75 & 12.41
	        \\
	        \hhline{|~|~|-||-|-|-|}
			&  & dEIRL & 4.85 & 10.28 & 7.98
	        \\
	        \hhline{|~|~|-||-|-|-|}
			&  & Opt. LQ & 4.65 & 10.01 & 7.59
	        \\	       	        
	        \hline		                	                
	\end{tabular}
	\end{minipage}	
	\label{tb:ES_CL_error_step_resp_characteristics}
\end{table}

\begin{figure}[h]
	\begin{center}
		\begin{tabular}{l}
			\includegraphics[height=\figsize]{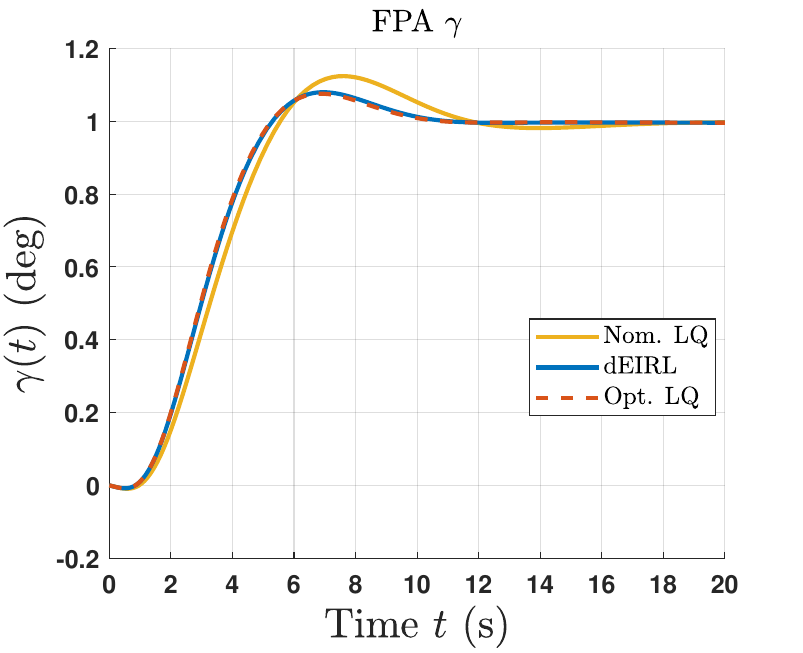}	
		\end{tabular}
	\end{center}
	\caption{Closed-loop 1$^{\circ}$ FPA step response for 25\% lift-coefficient modeling error $\nu = 0.75$ (\ref{eq:C_L_a}).}
	\label{fig:ES_CL_error_step_g}	
\end{figure}

As can be seen in Table \ref{tb:ES_CL_error_step_resp_characteristics}, dEIRL successfully recovers the closed-loop step response characteristics of the optimal LQ controllers. The performance recovery is most apparent in the FPA loop $j = 2$, where for the severe modeling error $\nu = 0.75$ the performance of the nominal LQ controller is significantly degraded compared to that of dEIRL and the optimal. In particular, the 1\% FPA settling time $t_{s, \gamma, 1 \%}$ of the nominal LQ controller is almost 17 s, and only 10 s for dEIRL and the optimal LQ. Similarly, percent overshoot in FPA  $M_{p,\gamma}$ is over 12\% for the nominal LQ controller, and only 8\% for dEIRL and the optimal LQ. The corresponding FPA step response for the 25\% lift-coefficient modeling error $\nu = 0.75$ is plotted in Figure \ref{fig:ES_CL_error_step_g}. 
Corroborated numerically by Table \ref{tb:ES_CL_error_step_resp_characteristics}, Figure \ref{fig:ES_CL_error_step_g} shows that dEIRL has qualitatively recovered the LQ-optimal closed-loop step response performance in spite of the severe 25\% modeling error.
As a brief aside, the first $t = 1$ s of the FPA response in Figure \ref{fig:ES_CL_error_step_g} exhibits the characteristic inverse nonminimum phase behavior attributable to the parasitic downward lift induced by pitch-up elevator deflections $\delta_{E}$.

%
%

\renewcommand{\relpathone}{nu_0p75/ES_CL_error_step_g/}	

\section{Conclusion \& Discussion}\label{sec:conclusion}

This work presents a suite of innovative CT-RL algorithms which rely on two design elements, multi-injection and decentralization, in order to address the performance limitations facing ADP-based CT-RL algorithms discovered in \cite{BA_Wallace_J_Si_CT_RL_review:2022}.
We develop these new algorithms with accompanying results on theoretical convergence and stability guarantees. Our in-depth quantitative performance evaluations of the four algorithms show that MI and decentralization (i.e., the dEIRL algorithm) achieve conditioning reductions of multiple orders of magnitude. Furthermore, evaluations show convergence and stability results corroborating theoretical analysis, and that the algorithms successfully recover the optimal controller and closed-loop performance in the presence of severe modeling errors. Altogether, our numerical studies demonstrate that dEIRL is a highly intuitive and effective design tool. This is a significant step forward from results of previous CT-RL algorithms, which ultimately fail to synthesize stabilizing designs even for simple academic systems \cite{BA_Wallace_J_Si_CT_RL_review:2022}.

We would like to point out that decentralization allows the designer to select learning parameters optimized to the inherent physics of each loop, rather than being forced to select a single set of “middle-ground” parameters which fails to adequately address individual-loop learning needs (cf. Section \ref{ssec:ES_setup}). Furthermore, weak dynamical coupling means that the excitation selections $d_{j}$, $r_{j}$ in one loop have little effect on data quality in other loops, and the rest of the hyperparameters have no inter-loop effects. This allows the designer to troubleshoot at the individual loop level, greatly simplifying the design process. With all the demonstrated benefits of decentralization, it is noted that this decentralization is physics-driven, with many important applications in areas such as robotics and aerial vehicles.



%
%





%
%


\bibliographystyle{IEEEtran}
\bibliography{\bibpath refs}



\vfill

\end{document}